\documentclass[prb,notitlepage,superscriptaddress]{revtex4-2}
\usepackage[T1]{fontenc}
\usepackage{textcomp}
\usepackage{amsmath}
\usepackage{amssymb}
\usepackage{graphicx}
\usepackage{esint}
\begin{document}
\title{Spatially strongly confined atomic excitation via two dimensional stimulated Raman adiabatic passage}

\author{Hamid R. Hamedi}
\email{hamid.hamedi@tfai.vu.lt}

\affiliation{Institute of Theoretical Physics and Astronomy, Vilnius University, Saul\.etekio 3, Vilnius LT-10257, Lithuania}

\author{Giedrius \v{Z}labys}

\affiliation{Institute of Theoretical Physics and Astronomy, Vilnius University, Saul\.etekio 3, Vilnius LT-10257, Lithuania}

\author{Ver\`onica Ahufinger}

\affiliation{ Departament de {F\'isica}, Universitat { Aut\`onoma} de Barcelona, E-08193 Bellaterra, Spain}

\author{Thomas Halfmann}

\affiliation{Institut fur Angewandte Physik, Technische Universitat Darmstadt, Hochschulstrasse 6, 64289 Darmstadt, Germany}
\author{Jordi Mompart}

\affiliation{ Departament de {F\'isica}, Universitat { Aut\`onoma} de Barcelona, E-08193 Bellaterra, Spain}
\author{Gediminas Juzeli\=unas}

\affiliation{Institute of Theoretical Physics and Astronomy, Vilnius University, Saul\.etekio 3, Vilnius LT-10257, Lithuania}
\begin{abstract}
We consider a method of sub-wavelength superlocalization and patterning of atomic matter waves via a two dimensional stimulated Raman adiabatic passage (2D STIRAP) process. An atom initially prepared in its ground level interacts with a doughnut-shaped optical vortex pump beam and a traveling wave Stokes laser beam with a constant (top-hat) intensity profile in space. The beams are sent in a counter-intuitive temporal sequence, in which the Stokes pulse precedes the pump pulse. The atoms interacting with both the traveling wave and the vortex beam are transferred to a final state through the 2D STIRAP, while those located at the core of the vortex beam remain in the initial state, creating a super-narrow nanometer scale atomic spot in the spatial distribution of ground state atoms. By numerical simulations we show that the 2D STIRAP approach outperforms the established method of coherent population trapping, yielding much stronger confinement of atomic excitation. Numerical simulations of the Gross-Pitaevskii equation show that using such a method one can create 2D bright and dark solitonic structures in trapped Bose-Einstein condensates (BECs). The method allows one to circumvent the restriction set by the diffraction limit inherent to conventional methods for formation of localized solitons, with a full control over the position and size of nanometer resolution defects.
\end{abstract}
\maketitle
\section{Introduction}
Diffraction usually imposes the limitation that the atomic position
cannot be detected more precisely than a half of the wavelength of
the radiation used for the detection. This restricts the possible
resolution, that is the minimal spatial extent of atomic excitations,
driven by focused laser beams. For instance, the diffraction limit
prohibits the high-fidelity control of individual atoms in a neutral-atom
quantum computing architecture, if they are separated by less
than the wavelength of light. In lens-based light microscopes, the
image of a point object obtained from the fluorescence emitted by
a fluorophore placed in the sample is, in principle, limited by diffraction.
Progress in the field will benefit from optical patterning and localization
of atomic matter waves with a resolution not limited by the wavelength
of the radiation involved. Some important examples are precise addressing
of single or few atoms in quantum memories (${\it e.g.}$ to generate
qubits) \cite{Monroe1995,Cirac1995,Isenhower2010,Wilk2010}, patterning
of Bose-Einstein condensates (e.g., for applications in quantum information
processing) \cite{Staliunas2002,Modugno2006}, high-resolution imaging
(e.g. labelling in two-photon fluorescence microscopy) \cite{Hell2007,Juan2013},
or optical lithography (${\it e.g.}$ by precisely localized excitations
in a photoresist) \cite{Thywissen2005}.

Modern optical imaging provides some ways to overcome the diffraction
limit. A very powerful approach for super-resolution microscopy is
based on stimulated emission depletion (STED) \cite{Hell2007,Hell2009}.
STED makes use of interactions, driven by two laser beams in an organic
fluorophore. However, interactions in STED are purely incoherent. On the other hand, while STED fluorescence microscopy allows theoretically to achieve a resolution at the size of a single fluorophore, in practice they are typically limited to about $\sim$20nm \cite{LocalizationSTED}. As an alternative approach, one can use coherent-adiabatic light-matter
interaction schemes to tightly localize excitation volumes \cite{PaspalakisPhysRevA.63.065802,SahraiPhysRevA.72.013820,Agarwal2006,Gediminas2007,GorshkovPhysRevLett.100.093005,ProitePhysRevA.83.041803,MilesPhysRevX.3.031014,Miles2015}.
The key ideas are to utilize the dark states in coherent population
trapping (CPT) \cite{Alzetta1976}, electromagnetically induced transparency
(EIT) \cite{Harris1997,Fleischhauer2005,Heinze2013}, or stimulated
Raman adiabatic passage (STIRAP) \cite{Bergmann1998,Ivanov2004,Vitanov2017,Klein2008}. It has been experimentally demonstrated that the CPT
configuration allows to produce fairly complex excitation profiles
with feature sizes as small as 60nm \cite{Miles2015}.
EIT and STIRAP make use of similar coupling schemes as STED. However,
in contrast to the conventional (incoherent) approaches, the schemes
involving coherent-adiabatic interaction converge much faster towards
high resolution and permit much stronger confinement of atoms. 

As an important feature of adiabatic interactions, there is a pronounced
nonlinear dependence of coherent excitation probabilities versus the
driving laser intensities. As an example, let us consider the adiabatic
population transfer by STIRAP. The process requires two strong laser
fields, i.e., a pump and a Stokes fields acting on atoms in a $\Lambda$-type
level scheme (see Fig. 1(a)). Let us assume that
initially all atomic population is in the ground state $|a\rangle$.
If we consider incoherent excitation with overlapping laser beams
(as in conventional stimulated emission pumping (SEP), which is similar
to the STED configuration), we expect that eventually an equal population
is established in all three atomic states. Hence, the population transfer
to the final state $|b\rangle$ would be neither complete nor selective
for the incoherent excitation. On the other hand, the coherent STIRAP
process involving a counter-intuitive pulse sequence (the Stokes pulse
preceding the pump pulse) enables a complete and selective population
transfer from the state $|a\rangle$ to the $|b\rangle$ via the adiabatic
following of the dark atomic dressed state during the atom-light interaction
process \cite{Bergmann1998,Ivanov2004,Vitanov2017,Klein2008}.

Adiabatic processes which require the driving laser intensities to
exceed an adiabaticity threshold, exhibit a very strong nonlinear
behaviour. Consequently, the spatial resolution of adiabatic excitation
increases rapidly (much faster compared to the case of the incoherent
excitation) with increasing the laser intensity, and is not limited
by the diffraction. Spatially confined atomic excitation based on
STIRAP is a fully coherent process that does not rely on the spontaneous
emission, as the atoms adiabatically follow the dark state \cite{Gediminas2007,MompartPhysRevA.79.053638,jordiOE}.
Therefore, it can be also important for systems, where coherence has
to be preserved, e.g., atomic Bose-Einstein condensates (BECs), or
for applications to the quantum information science. We note, that
while the STIRAP technique has already found a multitude of applications
\cite{Bergmann1998,Ivanov2004,Vitanov2017,Klein2008}, thus far it
has been rarely applied for high-precision atomic localization \cite{Gediminas2007,MompartPhysRevA.79.053638,jordiOE}.

The first theory proposals and experimental investigations on adiabatically
localized atomic excitations dealt with EIT (or CPT) \cite{Agarwal2006,PaspalakisPhysRevA.63.065802,SahraiPhysRevA.72.013820,ProitePhysRevA.83.041803,MilesPhysRevX.3.031014,Miles2015,GorshkovPhysRevLett.100.093005, local1,local2,local3}
rather than STIRAP \cite{Gediminas2007,MompartPhysRevA.79.053638}.
In EIT a dark dressed state emerges, which permits confinement of
atomic excitation \cite{Agarwal2006,GorshkovPhysRevLett.100.093005}.
In contrast to STIRAP with two strong and delayed pulses, EIT makes
use of overlapping pulses with a weak probe and a strong coupling
pulse. While STIRAP was developed for population transfer, EIT was
initially designed to reduce absorption of a weak probe beam by suppressing
excitation from the ground state coupled to the excited state by a
stronger control (coupling) field \cite{Harris1997,Fleischhauer2005}.
Note that in the case of the EIT the spatial confinement does not
converge as fast as for STIRAP. The earlier localization protocols
with standing waves create a periodic pattern of tightly localized
regions, which was fine for the first experimental demonstrations
\cite{ProitePhysRevA.83.041803,MilesPhysRevX.3.031014,Miles2015},
but it is not suitable for most applications which usually require
single excitation regions.

To obtain a single excitation region (spot localization), here we
combine the STIRAP with a STED-like beam geometry by expanding the
initial idea for the STIRAP-based localization approach \cite{Gediminas2007,MompartPhysRevA.79.053638}.
The atoms are initially prepared in their ground state and subsequently
adiabatically follow the dark state by applying the STIRAP scheme
\cite{Bergmann1998,Ivanov2004,Vitanov2017,Klein2008}, in which both
pulses act in a counter-intuitive temporal sequence, the Stokes pulse
preceding the pump one. In the following we assume a Stokes beam with a constant intensity across the interaction region, while the pump beam has a doughnut-like spatial profile. We also assume the latter to be an optical vortex carrying an orbital angular
momentum (OAM) \cite{Allen1999} (see Fig. 1(a)).
In that case, atoms remain in their initial ground state when located
at the vortex core (zero intensity) of the pump beam. On the other
hand, away from the vortex core the atoms interacting with both the
Stokes and the pump fields are transferred to the final state through
the 2D STIRAP process. The atoms localized at the pump core are not excited by light and therefore do not suffer from recoil-induced broadening, which implies that the Raman-Nath approximation perfectly applies (see \cite{Meystre2001} and references therein). On the other hand, the atoms that change their internal state as a consequence of the STIRAP process do not suffer from spontaneous emission since the process is coherent. For the latter, the total momentum exchanged corresponds only to the momentum difference between the doughnut field and the traveling wave field, which is almost negligible if the energies of two internal states are close \cite{MompartPhysRevA.79.053638}.

 It should be pointed out that a similar configuration has been employed in  Ref. \cite{jordiOE} with Stokes pulse having a Bessel beam spatial profile, and the one of the pump being the result of superimposing two Bessel beams focused with a lateral offset, producing a node in its center. On the other hand, in our scheme, the pump carries an OAM while the Stokes has no spatial dependence, creating a spot localization. The OAM of pump beam allows manipulation of the degree of localization and creates side spots in localization pattern, a feature which is missing in Ref. \cite{jordiOE}.

Using such a 2D STIRAP method, one can imprint topological phase singularities
on trapped atomic BECs, creating very narrow localized 2D density
defects. Specifically, the phase of the vortex beam is transferred
to the atoms during the STIRAP process, thus producing a 2D dark soliton
(a vortex) for the BEC atoms in another internal state. On the other
hand, the non-transferred atomic population makes a 2D bright soliton
located at the vortex core. These two component defects (bright and
dark solitons) are localized down to the order of a nanometer, thus
circumventing the restriction set by the diffraction limit. We have
carried numerical simulations of the Gross-Pitaevskii equation to
study these subwavelength structures. The method is also useful for
creating in a controllable way several dark and bright closely-spaced
solitons for studying their dynamics and collisional properties.

\section{Basic theory}
\begin{figure}
\includegraphics[width=0.5\columnwidth]{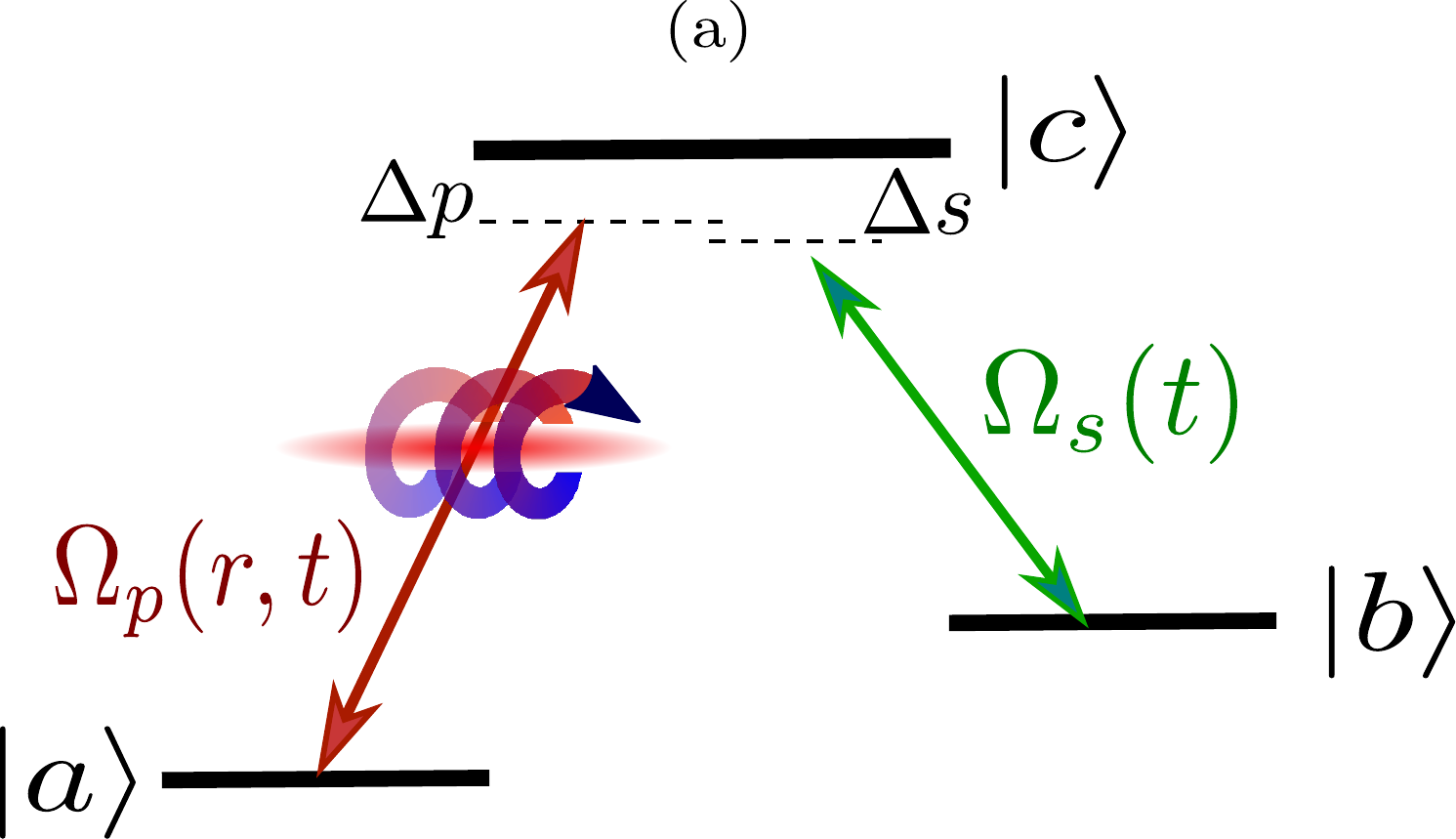}~~~~~~~~~~~~~~~~
\centering\includegraphics[width=0.3\columnwidth]{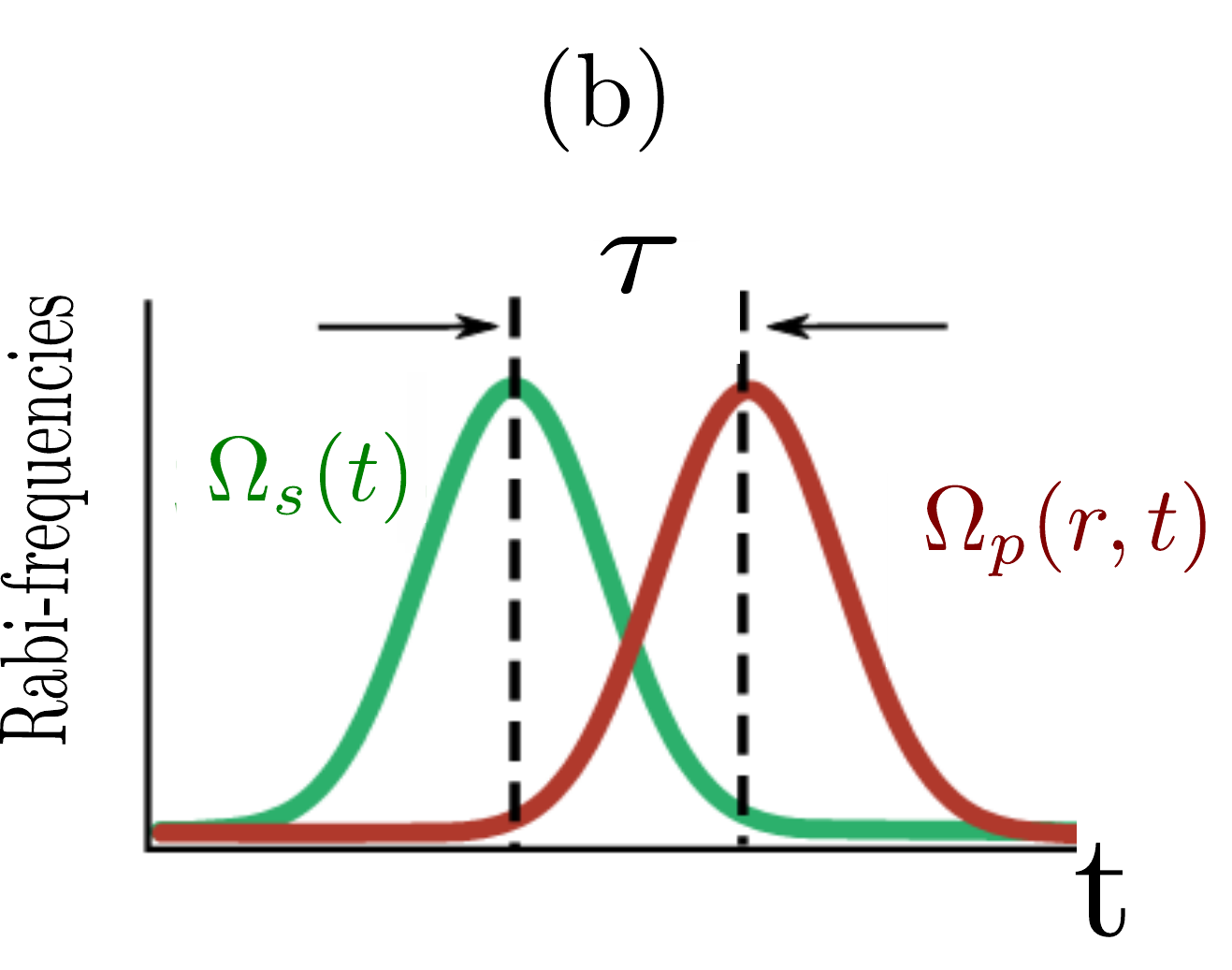} \caption{(a) Schematic representation of a three-level $\Lambda$-type atomic
system interacting with the pump and Stokes fields characterized by
the Rabi-frequencies $\Omega_{p}$ and $\Omega_{s}$, respectively.
(b) Temporal sequence of STIRAP pulses separated by the delay time
$\tau$.}
\label{fig:scheme-1}
\end{figure}

Let us consider an ensemble of atoms, characterized by a three-level
$\Lambda$-scheme of energy levels, as shown in Fig.~\ref{fig:scheme-1}(a).
The scheme includes an initial atomic state $|a\rangle$, an intermediate
excited level $|c\rangle$, and a target state $|b\rangle$. An atom
positioned at $\mathbf{r}$ interacts with a Stokes and a pump laser
pulses arriving with a time delay $\tau$ characterizing the STIRAP
process (Fig.~\ref{fig:scheme-1}(b)). The pulses are described by
the following slowly changing envelopes of Rabi frequencies 
\begin{equation}
\Omega_{s}(t)=\Omega_{s0}e^{-(t+\tau/2)^{2}/T^{2}},\label{eq:1-1}
\end{equation}
\begin{equation}
\Omega_{p}(\mathbf{r},t)=\Omega_{p}(\mathbf{r})e^{-(t-\tau/2)^{2}/T^{2}},\label{eq:2-1}
\end{equation}
where $T$ is a temporal width of the pulses. 

The Stokes laser field is a travelling wave acting on the transition
$|c\rangle\leftrightarrow|b\rangle$ with a constant spatial profile
$\Omega_{s0}$. On the other hand, the transition $|a\rangle\leftrightarrow|c\rangle$
is coupled by a doughnut-shaped Laguerre-Gaussian pump beam characterized
by the transverse profile 
\begin{equation}
\Omega_{p}(\mathbf{r})=\Omega_{p0}(\frac{r}{w})^{|l|}e^{-r^{2}/w^{2}}e^{il\phi},\label{eq:3-1}
\end{equation}
where $r$ describes the cylindrical radius, $w$ is the beam waist,
$\Omega_{p0}$ represents the strength of the doughnut beam and $\phi$
denotes the azimuthal angle with respect to the beam axis $z$. Such
a beam is characterized by the vorticity $l$ and carries an orbital
angular momentum (OAM) $\hbar l$ per photon \cite{Allen1999}.
The Stokes and pump detunings from the transitions $|c\rangle\leftrightarrow|b\rangle$ and $|a\rangle\leftrightarrow|c\rangle$ are denoted by $\Delta_{s}$
and $\Delta_{p}$, respectively. Assuming the one- and two-photon
resonance $\Delta_{s}=\Delta_{p}=0$, the quantum dynamics of the
$\Lambda$-type atom-light coupling can be described by the Hamiltonian
under the electric dipole and Rotating-wave-approximation (RWA)
\begin{equation}
H(t)=-\hbar\left[\Omega_{s}(t)|c\rangle\langle b|+\Omega_{p}(\mathbf{r},t),t)|c\rangle\langle a|\right]+H.c.\,.\label{eq:hamiltonian}
\end{equation}

The destructive interference of excitation pathways from $|a\rangle$
and $|b\rangle$ up to $|c\rangle$ ensures that a special superposition
of ground states 
\begin{equation}
|D(\mathbf{r},t)\rangle=\left[\Omega_{s}(t)|a\rangle-\Omega_{p}(\mathbf{r},t)|b\rangle\right]/\Omega(\mathbf{r},t)\label{eq:dark-1}
\end{equation}
known as the dark state is decoupled from both optical fields. The
dark state is an eigenstate of the atom-light Hamiltonian (\ref{eq:hamiltonian})
with a zero eigen-energy \cite{scully}. Here
\[
\Omega\equiv\Omega(\mathbf{r},t)=\sqrt{\left|\Omega_{s}(t)\right|^{2}+\left|\Omega_{p}(\mathbf{r},t)\right|^{2}}
\]
is the total Rabi frequency. The dark state can be represented as
\begin{equation}
|D(\mathbf{r},t)\rangle=\cos\,\theta(r,t)|a\rangle-\sin\,\theta(r,t)|b\rangle e^{il\phi},\label{eq:dark}
\end{equation}
where
\begin{equation}
\theta(r,t)=\arctan\left|\frac{\Omega_{p}(\mathbf{r})}{\Omega_{s0}}\right|e^{2\tau t/T^{2}}\,\label{eq:mixangle}
\end{equation}
 is the mixing angle.

\section{2D STIRAP superlocalization method}
The dark state (\ref{eq:dark-1}) does not include the leaky excited
state $|c\rangle$ decaying with a rate $\Gamma$. Therefore, the
losses will be suppressed if the atomic system adiabatically follows
the dark state $|D(\mathbf{r},t)\rangle$. The present 2D STIRAP superlocalization
protocol makes use of such an adiabatic following of the dark state
$|D(\mathbf{r},t)\rangle$ for the counterintuitive ordering of the
laser pulses shown in Fig. \ref{fig:scheme-1}(b), with the Stokes
pulse $\Omega_{s}(t)$ preceding the pump $\Omega_{p}(\mathbf{r},t)$ \cite{Bergmann1998,BergmannRevModPhys.70.1003},
so that   
\begin{equation}
\begin{array}{c}
\underset{t\rightarrow-\infty}{\lim}\frac{\Omega_{p}(\mathbf{r},t)}{\Omega_{s}(t)}=0,\\
\underset{t\rightarrow+\infty}{\lim}\frac{\Omega_{p}(\mathbf{r},t)}{\Omega_{s}(t)}=\infty.
\end{array}\label{eq:lim}
\end{equation}
 This implies that 
\begin{equation}
\begin{array}{c}
\theta(\mathbf{r},-\infty)=0,\\
\theta(\mathbf{r},+\infty)=\frac{\pi}{2}.
\end{array}\label{eq:teta}
\end{equation}
For such a pulse sequence, the dark state $|D(\mathbf{r},t)\rangle$
reduces to $|a\rangle$ for $t\rightarrow-\infty$ and to $|b\rangle e^{il\phi}$
for $t\rightarrow+\infty$. This provides a suitable vehicle for transferring
atoms from the initial state $|a\rangle$ to the final state $|b\rangle$
without populating the state $|c\rangle$. In order to maintain the
system in the dark state, the process should be carried out satisfying
a global adiabaticity condition \cite{BergmannRevModPhys.70.1003,VitanovRevModPhys.89.015006,Shore2017,MompartPhysRevA.79.053638}

\begin{equation}
\Omega_{p0}^{2}(\frac{r}{w})^{2|l|}e^{-2r^{2}/w^{2}}+\Omega_{s0}^{2}>(\frac{\beta}{\tau})^{2},\label{eq:adiabaticity}
\end{equation}
where $\beta$ is a dimensionless constant. For the optimal Gaussian
profiles and overlapping times, the parameter $\beta$ takes values
around $10$ \cite{KuklinskiPhysRevA.40.6741,MompartPhysRevA.79.053638}.

In this way, the atoms are initially in the internal state $|a\rangle$;
the Stokes field $\Omega_{s}(t)$, which couples initially unpopulated
states $|c\rangle$ and $|b\rangle$, is sent before the vortex pump
beam $\Omega_{p}(\mathbf{r},t)$, the latter coupling the populated
ground state $|a\rangle$ with the excited state $|c\rangle$. As
illustrated in Fig. \ref{fig:scheme-1}(b)), the two pulses should
overlap, so that the population could be transferred from the initial
state $|a\rangle$ to the final state $|b\rangle$ during the STIRAP.
In fact, if the temporal separation between the pulses $\tau$ is
too large, the adiabaticity condition (\ref{eq:adiabaticity}) fails
to hold. The intensity of the pump beam $\Omega_{p}(\mathbf{r},t)$
goes to zero at the vortex core, preventing the atoms situated in
this spatial region to be transferred to the final state $|b\rangle$.
On the other hand, away from the vortex core, the atoms interacting
with both the pump $\Omega_{p}(\mathbf{r},t)$ and Stokes $\Omega_{s}(t)$
fields are transferred to state $|b\rangle$ through the 2D STIRAP
process with a sufficiently large efficiency, see Appendix A. Therefore, if the amplitude of the vortex beam $\Omega_{p0}$
is considerably larger than the amplitude of the Stokes beam $\Omega_{s0}$,
a spot-like super-narrow spatial pattern can be generated at the core
of a strong pump beam when measuring the population distribution of
the atoms in the state $|a\rangle$.

\begin{figure}
\centering\includegraphics[width=0.8\columnwidth]{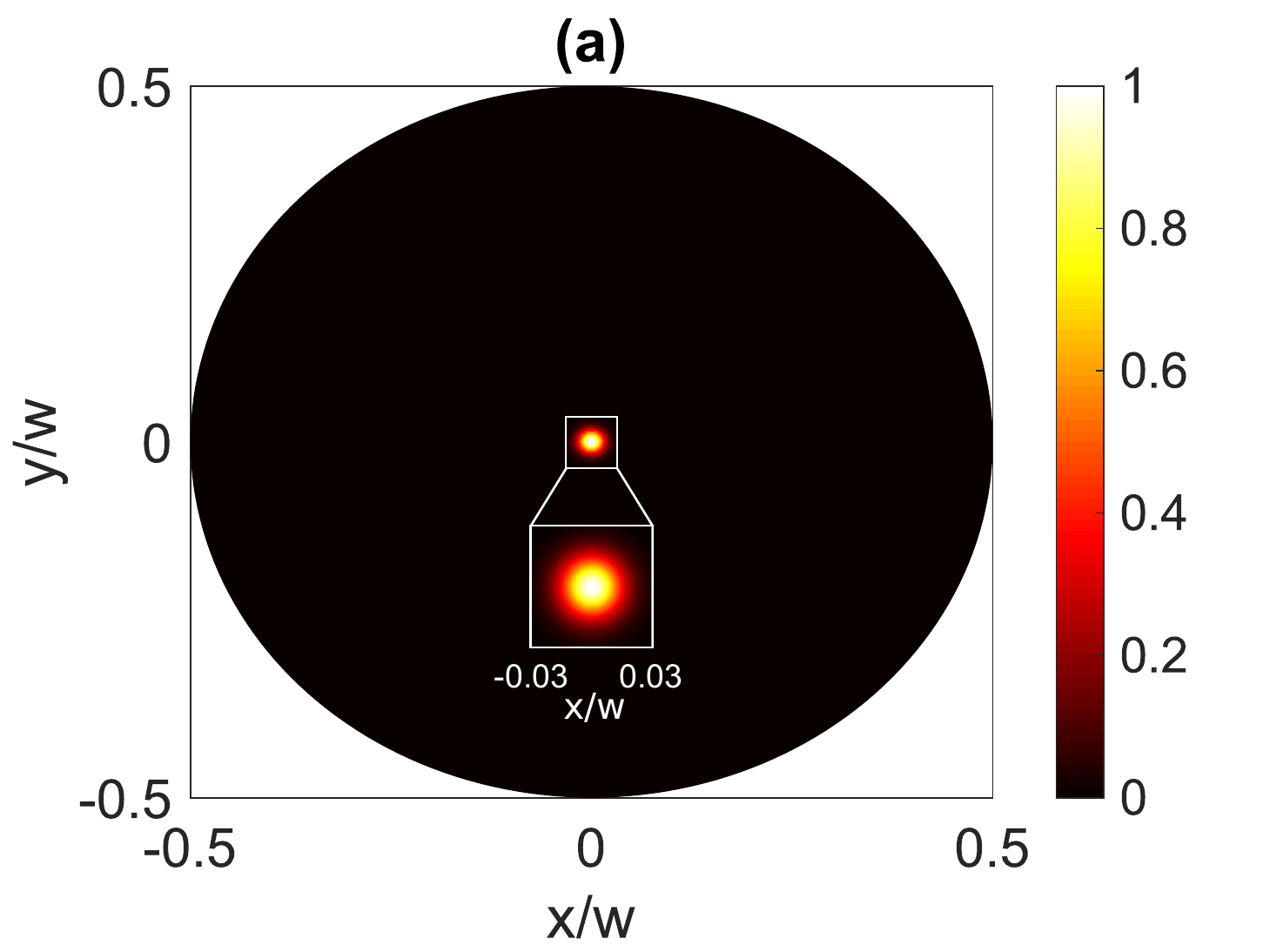}
\includegraphics[width=0.8\columnwidth]{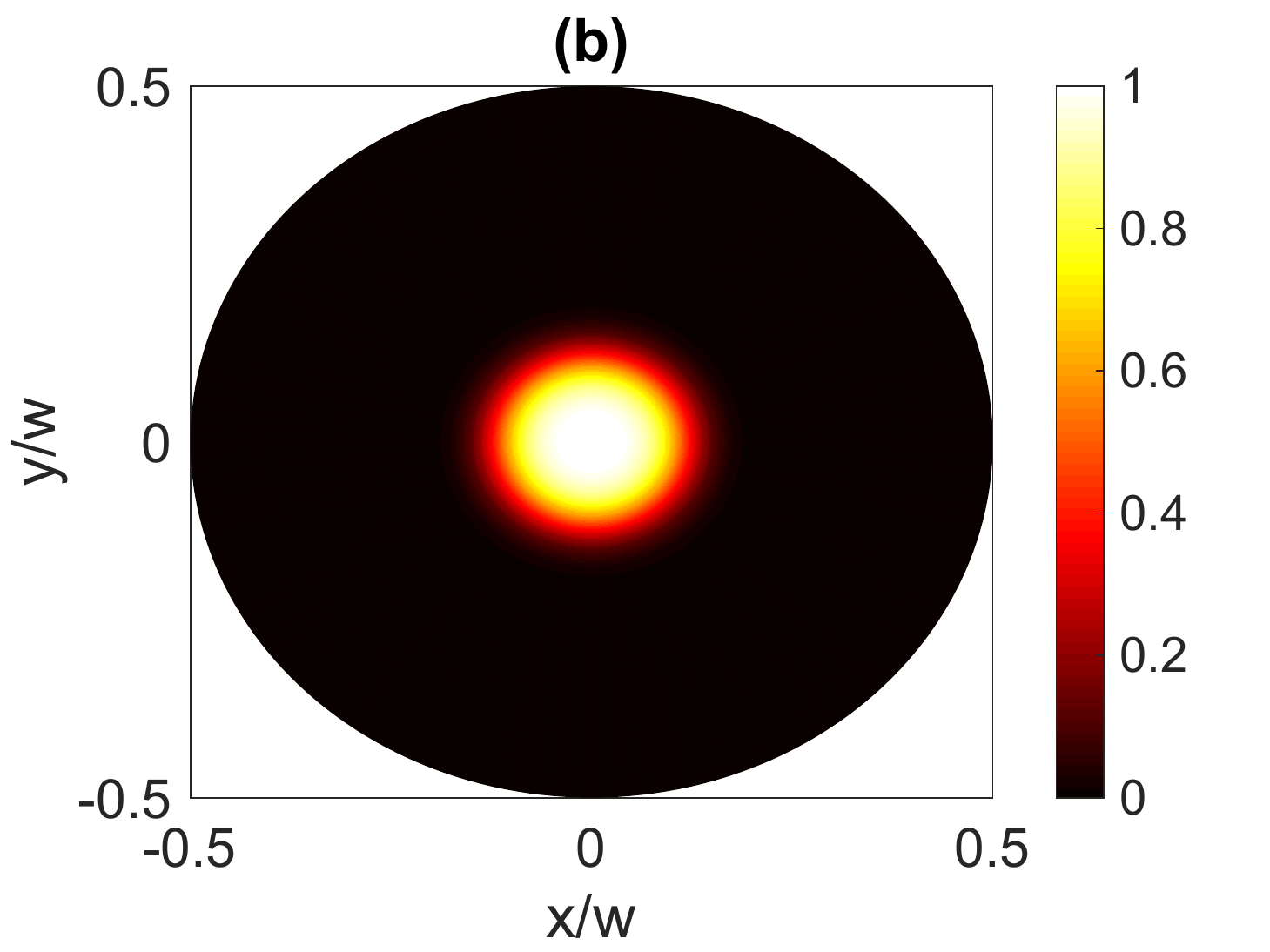}
\caption{Simulation of the spatial distribution of population of the state
$|a\rangle$ after applying the 2D STIRAP sequence. Here, $\Omega_{s0}=4\Gamma$,
$\mathcal{\alpha}=\Omega_{p0}^{2}/\Omega_{s0}^{2}=100$, $\tau=10/\Gamma$,
$T=5/\Gamma$ and (a) $l=1$, (b) $l=2$. The spatial coordinates
are normalized to the beam waist $w$. The color code indicates the
amount of population from $0(\%$ (Black) to $100\%$ (White). For
a beam waist of $w=1\mu m$, we confine the remaining population of
the state $|a\rangle$ in a narrow spot down to the range of a nanometer.}
\label{fig:fig2}
\end{figure}

Figure.~\ref{fig:fig2} depicts the population distribution of the
internal state $|a\rangle$ after applying the 2D STIRAP. The plots
are obtained by numerically solving the corresponding density matrix
equations of the $\Lambda$ system \cite{AgarwalPhysRevA.52.3147}
for $\mathcal{\alpha}=\Omega_{p0}^{2}/\Omega_{s0}^{2}=100$ and different
OAM numbers $l=1$ (a) and $l=2$ (b). For $l=1$ the atomic excitation
is seen to be strongly concentrated at the zero intensity region of
the pump field (at the core of vortex beam at $r\rightarrow0$), indicating
that the atoms can be well localized in a very narrow spot, much smaller
than the beam waist $w$ (Fig.~\ref{fig:fig2} (a)). From the inset in Fig.~\ref{fig:fig2} (a) we deduce a resolution (full width at half maximum, FWHM) along the x direction to be 0.02 in units of the beam waist for $\mathcal{\alpha}=\Omega_{p0}^{2}/\Omega_{s0}^{2}=100$. By taking the beam waist of $w=1\mu$m, we get a localization of $\sim$20nm, i.e. far below the diffraction limit. We checked by another numerical simulations (not shown in the paper) that the
resolution reduces to about $\sim$8nm for $\mathcal{\alpha}=1000$, and further decreases for larger values of   $\mathcal{\alpha}$.
 
When the OAM
number $l$ increases to $l=2$, the spot width increases resulting
in a lower localization precision, as can be seen in Fig.~\ref{fig:fig2}
(b). This is due to an increase of the core of the optical vortex
for the laser beams with a larger vorticity.
We also checked our results with a
Stokes beam of Gaussian intensity profile in space. The numerical simulation confirms, that the Stokes beam profile does not matter, provided the adiabaticity condition is still satisfied.

It is worthwhile clarifying that an optical vortex can be generated by an
intensity pattern with a Gaussian envelope that possesses a
point singularity at the center of the optical vortex beam \cite{GaussianVortex}.
This can be done by implementing a $4f$ lens arrangement to image
the optical vortex intensity profile immediately after a spiral
phase plate (SPP). However, such a vortex beam is not suitable for the purpose of atom localization. It is the  doughnut shaped spatial profile of the pump which creates the spot localization, as those atoms
located at the core of the pump beam (hollow core in the doughnut beam) remain in the initial state, creating the spot defects. The phase of the 
vortex beam in Eq.~(3) only controls the resolution of localization, as shown in (Fig.~\ref{fig:fig2} (a,b)).
\subsection{STIRAP vs. CPT for spot localization}

One can also achieve a spot-like localization in the three-level scheme
illustrated in Fig. \ref{fig:scheme-1}(a) by applying a CPT technique.
In this case, a spot localization is possible when the system reaches
the steady state through an optical-pumping process to the dark state
involving several cycles of laser excitation and the spontaneous emission.
The 2D STIRAP localization protocol described  previously
reduces to the 2D CPT case for the overlapping pump and Stokes pulses
($\tau=0$) with $\Gamma T\gg1$. In this case, an atom initially
prepared in the state $|a\rangle$ will end up in the dark state at
the steady state if the two-photon resonance is maintained. As the
intensity of the pump beam goes to zero at the vortex core, the dark
state of the system reduces to the bare state $|a\rangle$ there,
as one can see in Eqs.~(\ref{eq:dark-1}) and (\ref{eq:dark}). Hence
the population distribution of the state $|a\rangle$ shows a maximum
at the center of  the vortex beam.

\begin{figure}
\centering\includegraphics[width=0.8\columnwidth]{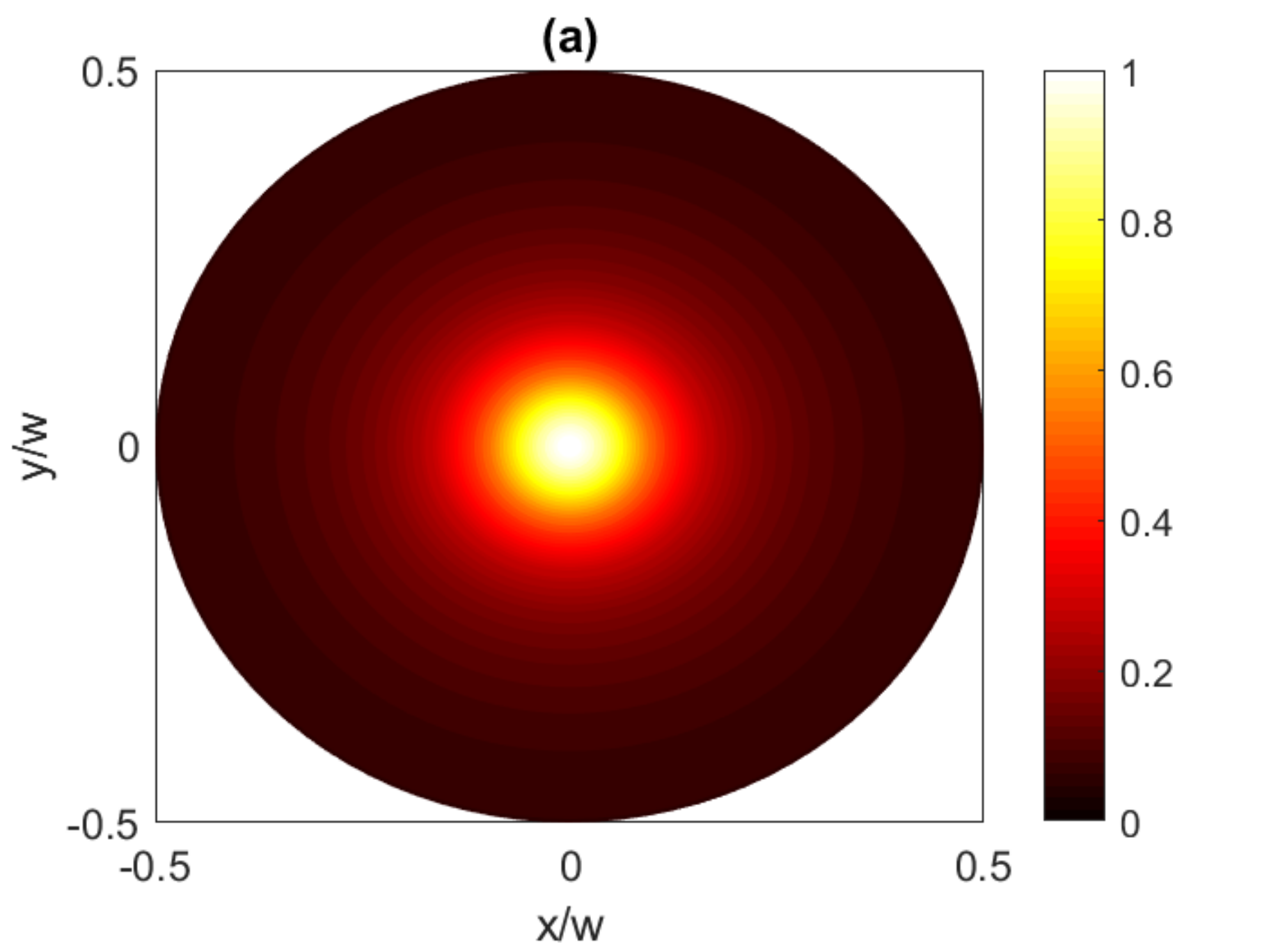} 
\includegraphics[width=0.8\columnwidth]{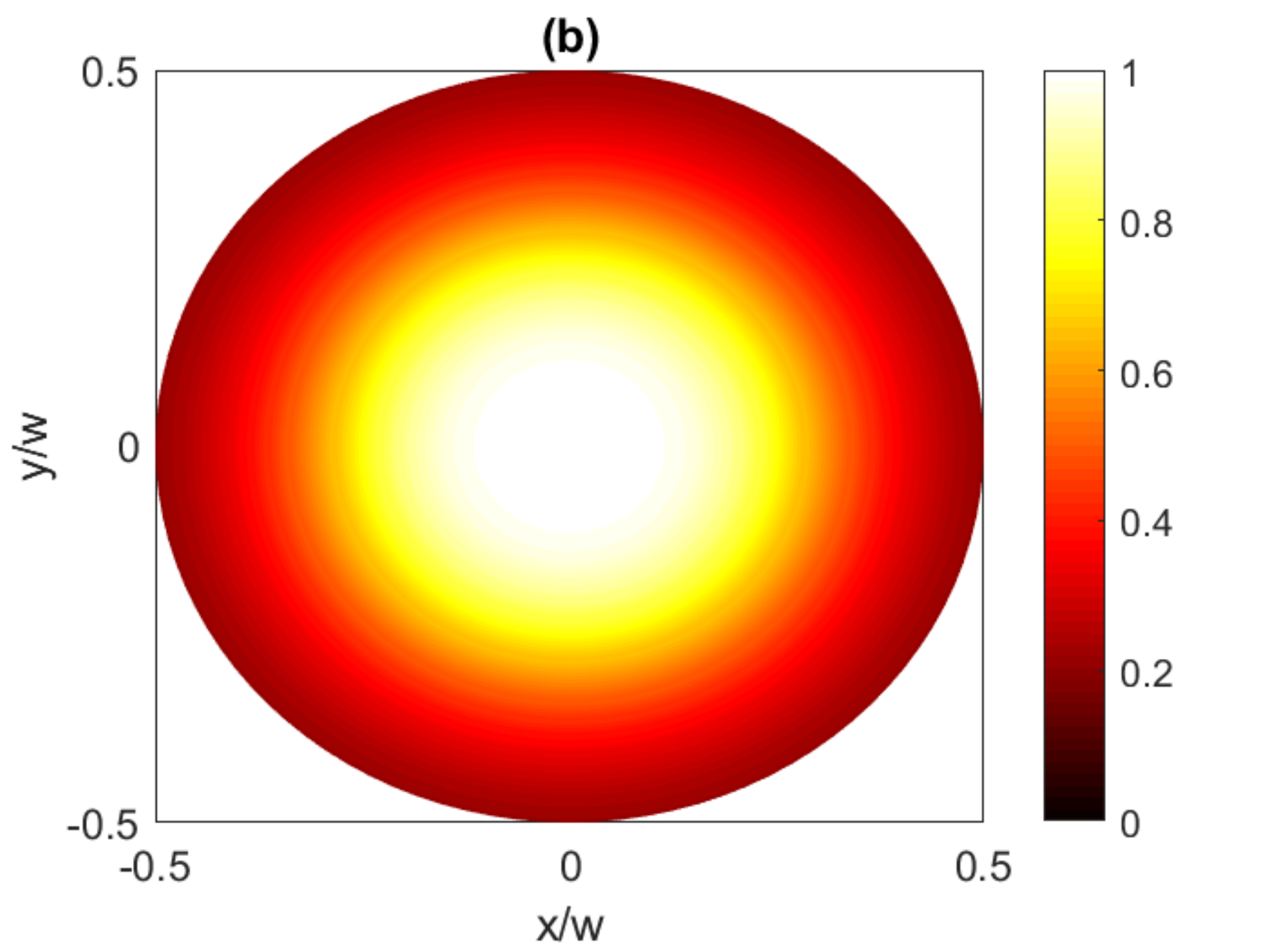}

\caption{Simulation of the spatial distribution of the population in the state
$|a\rangle$ after applying the CPT technique and for (a) $l=1$ and
(b) $l=2$. Here, the parameter values are the same as in Fig. \ref{fig:fig2},
except that there is no delay time between the two pulses $\tau=0$.
Spatial coordinates are normalized to the beam waist $w$. The color
code indicates the amount of population from $0(\%$ (Black) to $100\%$
(White).}
\label{fig:fig3-1}
\end{figure}

In Fig.~\ref{fig:fig3-1} we present the population of the state
$|a\rangle$ with different OAM numbers $l=1$ (a) and $l=2$ (b)
after applying the 2D CPT protocol. The population distribution shows
again a spot-like pattern at the center. One can see that the localization
patterns can be well controlled by the OAM number $l$, like in the
case of the 2D STIRAP presented in Fig.~\ref{fig:fig2}. In both
cases the narrowest spot structure could be always achieved with the
minimum winding number ($l=1$) corresponding to the smallest core
of the optical vortex. Yet, the localization precision is lower for
the 2D CPT, indicating that the 2D STIRAP permits much stronger 2D
confinement. 

\subsection{Off-axis spot localization}

\begin{figure}
\centering\includegraphics[width=0.55\columnwidth]{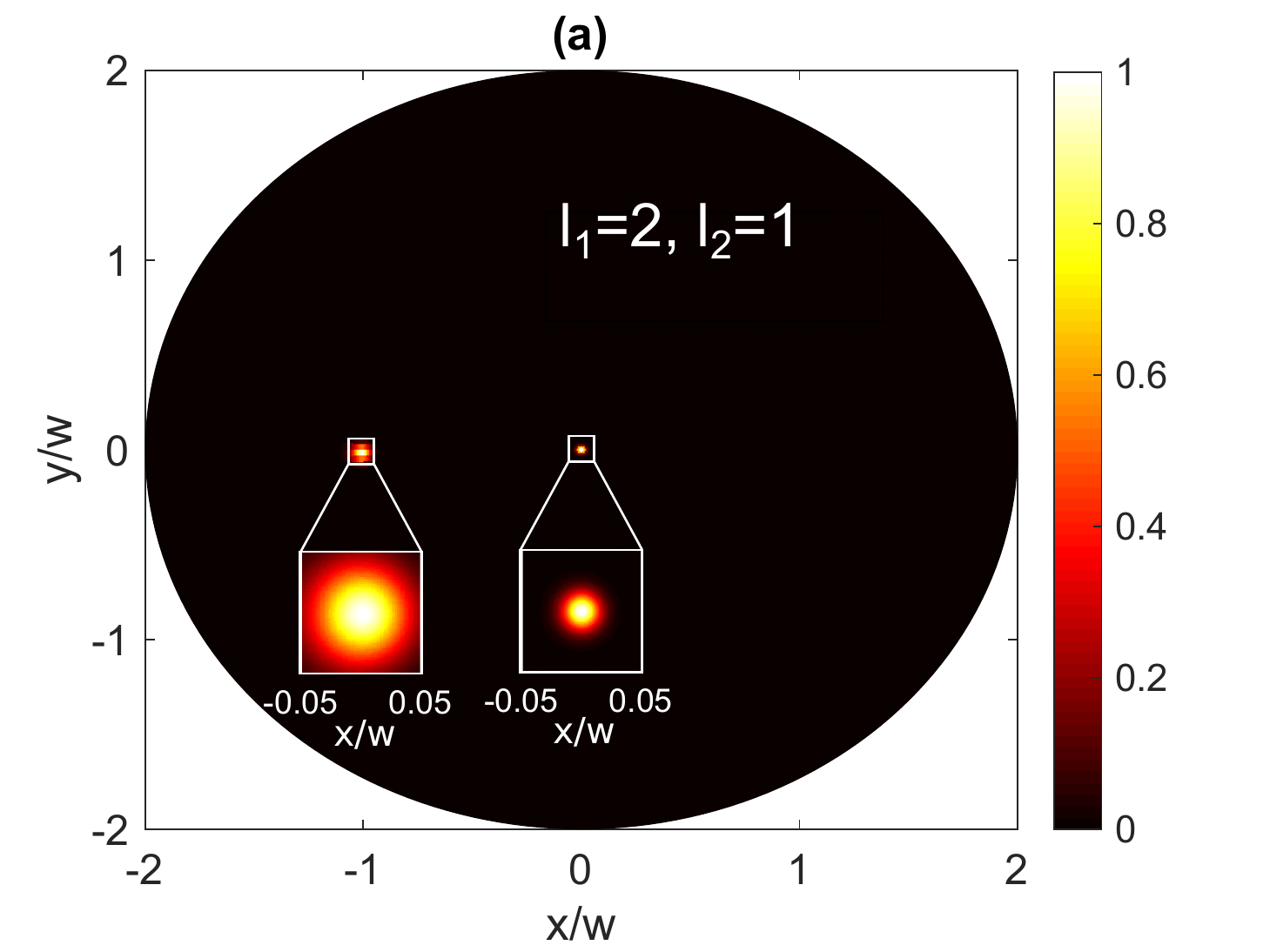}\includegraphics[width=0.55\columnwidth]{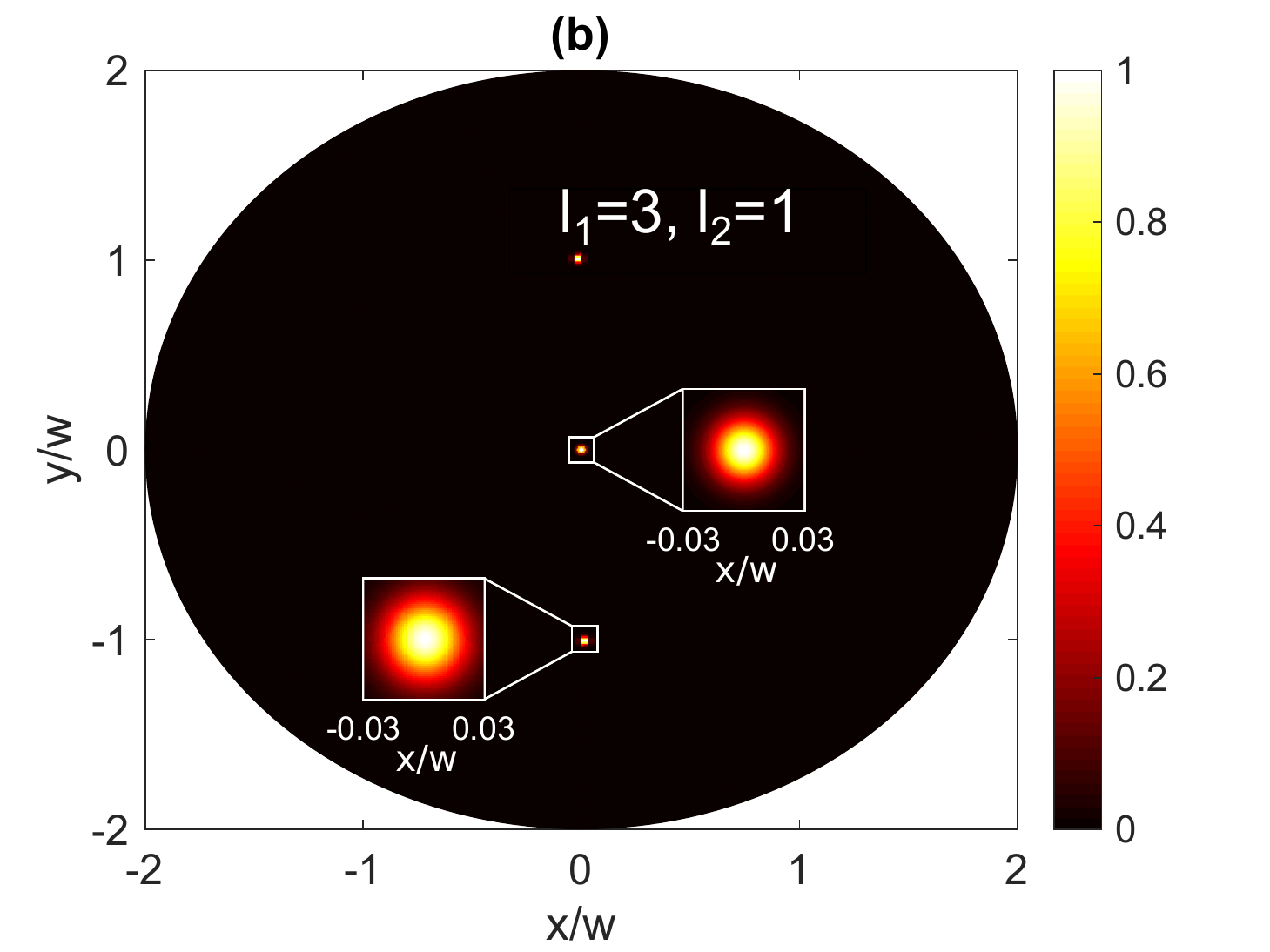}~~\\
\includegraphics[width=0.55\columnwidth]{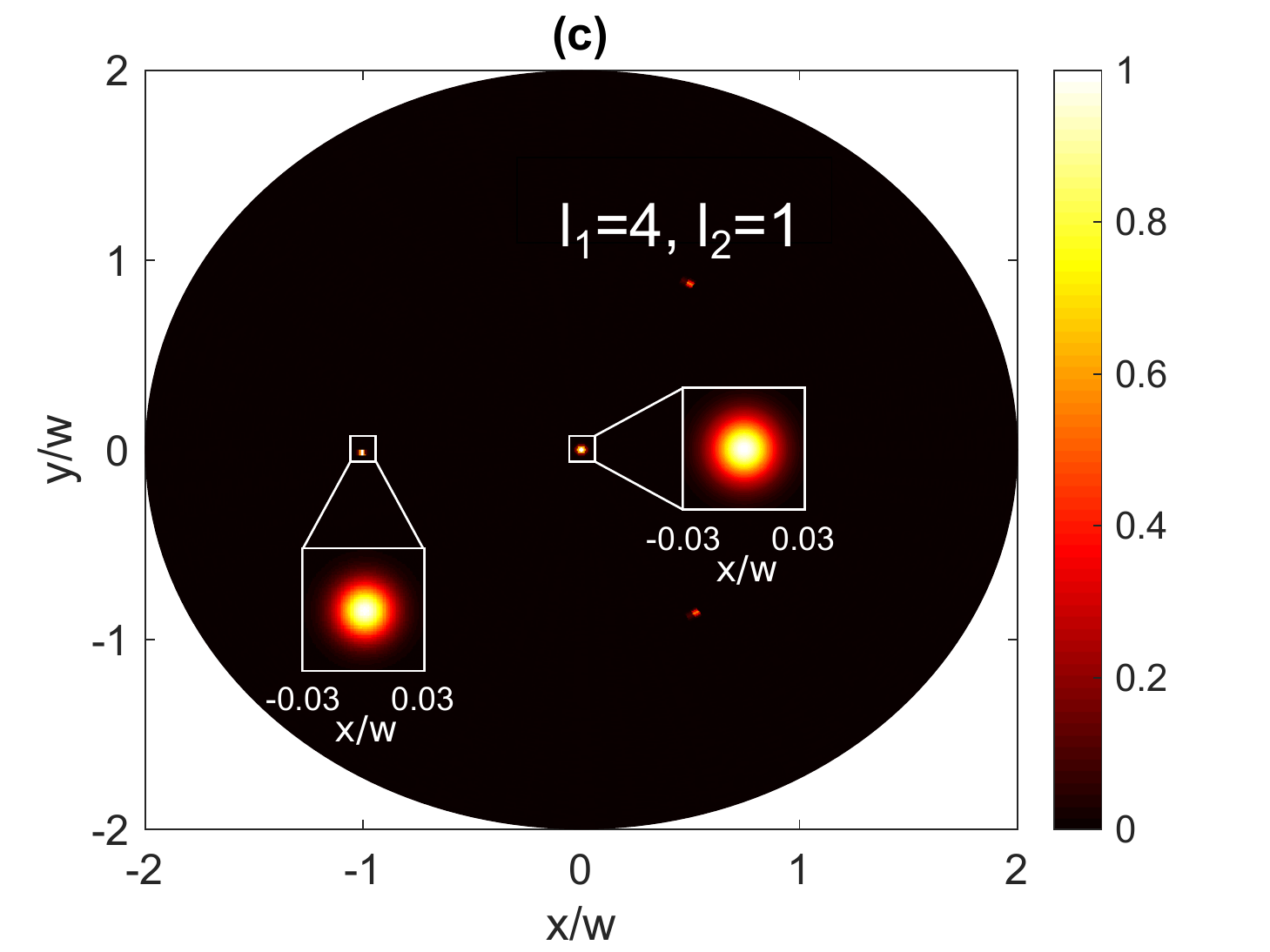}\includegraphics[width=0.55\columnwidth]{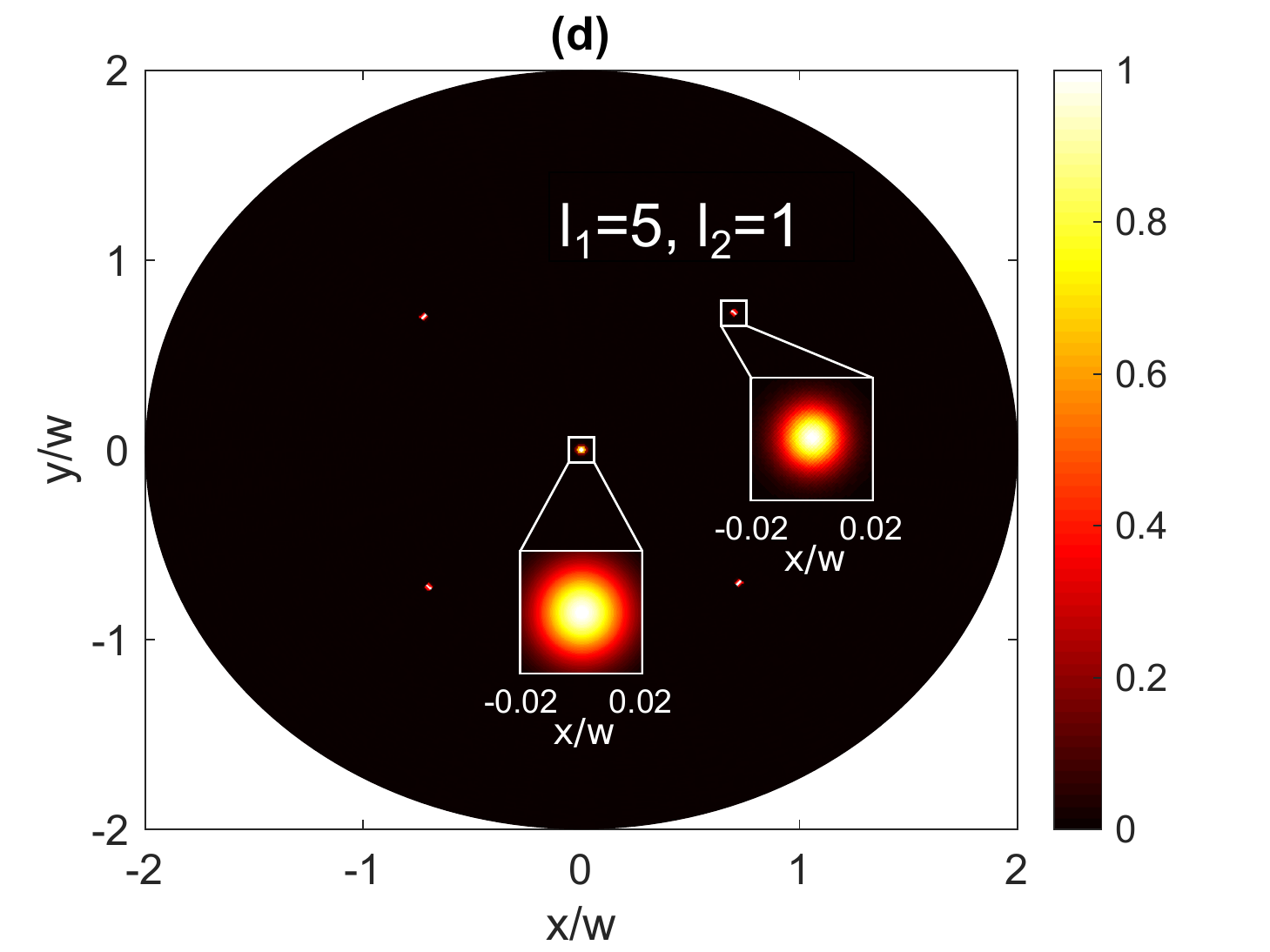}

\caption{Simulation of the spatial distribution of population in the state
$|a\rangle$ for the composite beam described by Eq.~(\ref{eq:composite beam})
with different OAM numbers $l_{1}$ and $l_{2}$ after applying the
STIRAP sequence. Here, $\Omega_{p_{1}0}=\Omega_{p_{2}0}=\Omega_{p0}$
and the parameter values are the same as Fig. \ref{fig:fig2}. Spatial
coordinates are normalized to the beam waist $w$. The color code
indicates the amount of population from $0(\%$ (Black) to $100\%$
(White).}
\label{fig:fig4}
\end{figure}

When two optical beams with different OAM numbers ($|l_{1}|<|l_{2}|$)
are superimposed collinearly they generate a pattern of vortices that
depends on the relative amplitude and phase of individual beams \cite{Franke-Arnold2007,Baumann2009}.
We consider now such a superposition of two collinear component beams
with topological charges $l_{1}$ and $l_{2}$ for the purpose of
off-axis 2D STIRAP localization. In this case, the pump beam reads 
\begin{equation}
\Omega_{p}(\mathbf{r},t)=\Omega_{p_{1}0}(\frac{r_{1}}{w})^{|l_{1}|}e^{-r_{1}^{2}/w^{2}}e^{il_{1}\phi_{1}}e^{-(t-t_{p})^{2}/T^{2}}+X\Omega_{p_{2}0}(\frac{r_{2}}{w})^{|l_{2}|}e^{-r_{2}^{2}/w^{2}}e^{il_{2}\phi_{2}}e^{-(t-t_{p})^{2}/T^{2}},\label{eq:composite beam}
\end{equation}
where $X=1$, $r_{1}=r_{2}=r$ and $\phi_{1}=\phi_{2}=\phi$. The resulting
composite beam contains a vortex of charge $|l_{1}|$ located at the
beam center which is surrounded by $|l_{1}-l_{2}|$ peripheral vortices
\cite{Franke-Arnold2007,Baumann2009}. Figure \ref{fig:fig4} shows
the simulation of the spatial distribution of population in the state
$|a\rangle$ after applying the 2D STIRAP protocol and using the composite
pump beam given by Eq.~(\ref{eq:composite beam}). One can see that
the spatial profile of the population develops new off-axis atomic
spots depending on the topological charges $l_{1}$ and $l_{2}$,
which enables single-site addressability of trapped arrays of atoms.
These peripheral localization spots are evenly distributed at angles
$\phi_{L}=\frac{n\pi}{\Delta l},$ where $\Delta l=l_{2}-l_{1}$ and
$n=1...(2|l_{2}-l_{1}|-1)$ is an odd integer for each of the peripheral
vortices.

\section{Localized 2D bright and dark solitons in a trapped BEC \label{sec:vort-dep-susc}}

We now focus on a trapped BEC of $^{87}Rb$ to show a feasibility
of generation of narrow structures in the condensate. We consider
a zero temperature two-species $^{87}Rb$ BEC, with $|a\rangle=|F=1,m_{f}=-1\rangle$
and $|b\rangle=|F=2,m_{f}=1\rangle$. To study vortex imprinting in
the BEC, the system is described by the 2D Gross-Pitaevskii equations
(GPEs) for a three-component BEC involving also an intermediate excited
state $|c\rangle$ needed to perform the STIRAP. Yet the excited state
is weakly populated if the adiabaticity condition (\ref{eq:adiabaticity})
holds.

Before the Stokes and pump laser fields are applied, the BEC is prepared
in the ground state $|a\rangle$ with the wave-function obeying the
stationary 2D GPE
\begin{equation}
\mu\psi_{a}=\left[-\frac{\hbar^{2}}{2m}\nabla^{2}+V_{a}(r)+Ng_{aa}|\psi_{a}|^{2}\right]\psi_{a}\,,\label{eq:BEC}
\end{equation}
with $\nabla^{2}=\left(\partial/\partial_{x}\right)^{2}+\left(\partial/\partial_{y}\right)^{2}$,
where $m$ is the atomic mass, $N$ is the number of atoms in the
condensate, $\mu$ indicates the chemical potential, $V_{a}(r)=\frac{m\omega_{r}^{2}r^{2}}{2}$
is the harmonic trapping potential, $r$ is the cylindrical radius,
$\omega_{r}$ is the angular frequency of the radial trap, $g_{ij}=\frac{\sqrt{8\pi}\hbar^{2}a_{ij}}{ma_{\perp}}$
describes the atom-atom interaction, $a_{ij}$ is the $s$-wave scattering
length between the species $i$ and $j$, and $a_{\perp}=\sqrt{\frac{\hbar}{m\omega_{\perp}}}$
denotes the characteristic length of the transverse harmonic trap
of the $2D$ like pancake shape BEC with $\omega_{\perp}\gg\omega_{r}$.
The parameter values used in the simulations are $m=1.44\times10^{-25}kg$,
$\omega_{r}=2\pi\times14\,s^{-1}$, $\omega_{\perp}=2\pi\times715\,s^{-1}$,
$a_{aa}:a_{ab}:a_{bb}=(1.03:1:0.97)\times55\mathring{A}$ and $N=5\times10^{5}$
\cite{MompartPhysRevA.79.053638}.

Subsequently the 2D multi-component BEC interacts with two laser pulses
characterized by Rabi-frequencies $\Omega_{p}$ and $\Omega_{s}$.
The atomic ground states described by wavefunctions $\psi_{a}$ and
$\psi_{b}$ are coupled to the intermediate excited state described
by the wavefunction $\psi_{c}$. The GPEs for evolution of such a
three-component BEC can be written as
\begin{equation}
\begin{array}{cc}
i\hbar\frac{d\psi_{a}}{dt}= & \left[-\frac{\hbar^{2}}{2m}\nabla^{2}+V_{a}(r)+Ng_{aa}|\psi_{a}|^{2}+Ng_{ab}|\psi_{b}|^{2}\right]\psi_{a}+\frac{1}{2}\hbar\Omega_{\mathrm{p}}(t,r,\phi)\psi_{c},\\
i\hbar\frac{d\psi_{b}}{dt}= & \left[-\frac{\hbar^{2}}{2m}\nabla^{2}+V_{b}(r)+Ng_{bb}|\psi_{b}|^{2}+Ng_{ab}|\psi_{a}|^{2}\right]\psi_{b}+\frac{1}{2}\hbar\Omega_{\mathrm{s}}(t)\psi_{c}\\
i\hbar\frac{d\psi_{c}}{dt}= & \frac{1}{2}\hbar\Omega_{\mathrm{p}}^{*}(t,r,\phi)\psi_{a}+\frac{1}{2}\hbar\Omega_{\mathrm{s}}(t)\psi_{b}-i\hbar\frac{\Gamma}{2}\psi_{c},
\end{array},\label{eq:GPEs}
\end{equation}
with the initial conditions $\psi_{a}(t=0)=\psi_{a}^{\mathrm{GS}}$
and $\psi_{b,c}(t=0)=0$. Here the trapping potential is the same
for both ground states  i.e., $V_{a}(r)=V_{b}(r)$, the Rabi frequencies
for the traveling wave $\Omega_{\mathrm{s}}(t)$ and the vortex beam
$\Omega_{p}(t,r,\varphi)$ are given by Eqs.~(\ref{eq:1-1}) and
(\ref{eq:2-1}), and we have omitted the kinetic energy and trapping
potential in the equations for the excited state wave-function. The
parameter values used for the time evolution simulation are $\mathcal{\alpha}=100$,
$l=1$, $\Omega_{s0}=2\pi\times10^{7}\,s^{-1}$, $\Omega_{p0}=\Omega_{s0}\sqrt{\mathcal{\mathcal{\alpha}}}$,
$T=5\mu s$, $t_{s}=20\mu s$, $t_{p}=30\mu s$, $\Delta_{p}=\Delta_{s}=0$,
$w=20\mu m$. The excited state decay is $\Gamma=2\pi\times5.41\times10^{6}\,s^{-1}$.
The space is discretized by a $513\times513$ grid, ranging from $-50\ \mu m$
to $50\ \mu m$ in both directions, hence the minimal discernible
distance is $100/513\ \mu m$. The system is evolved for $t=100\ \mu s$.
The simulations are performed using the GPELab toolbox\cite{Antoine2014Nov,Antoine2015Aug}. 

\begin{figure}
\centering\includegraphics[width=0.6\columnwidth]{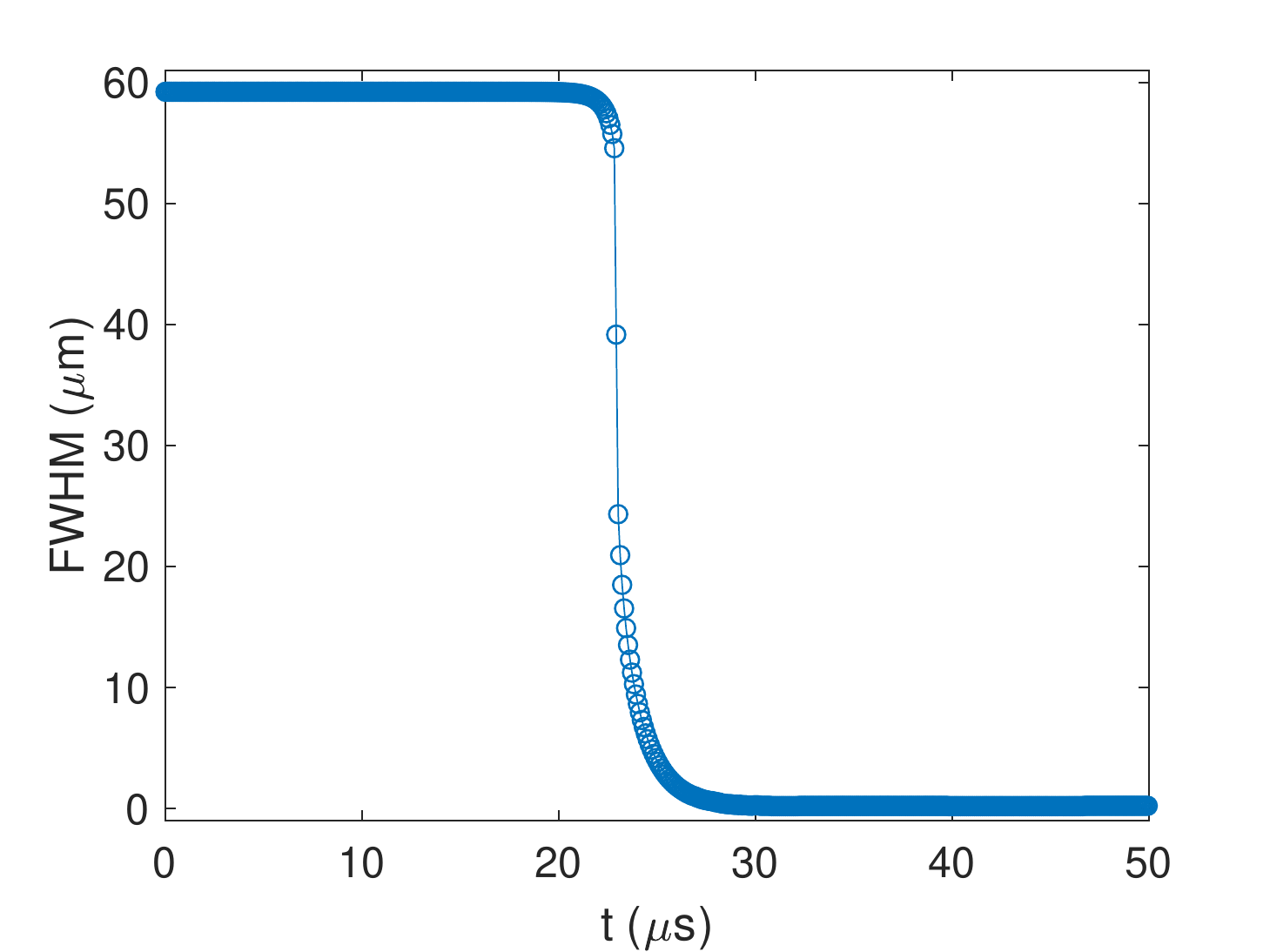}

\caption{Time evolution of the FWHM of density of component $a$.}
\label{fig:fig6}
\end{figure}

\begin{figure}
\centering\includegraphics[width=0.5\columnwidth]{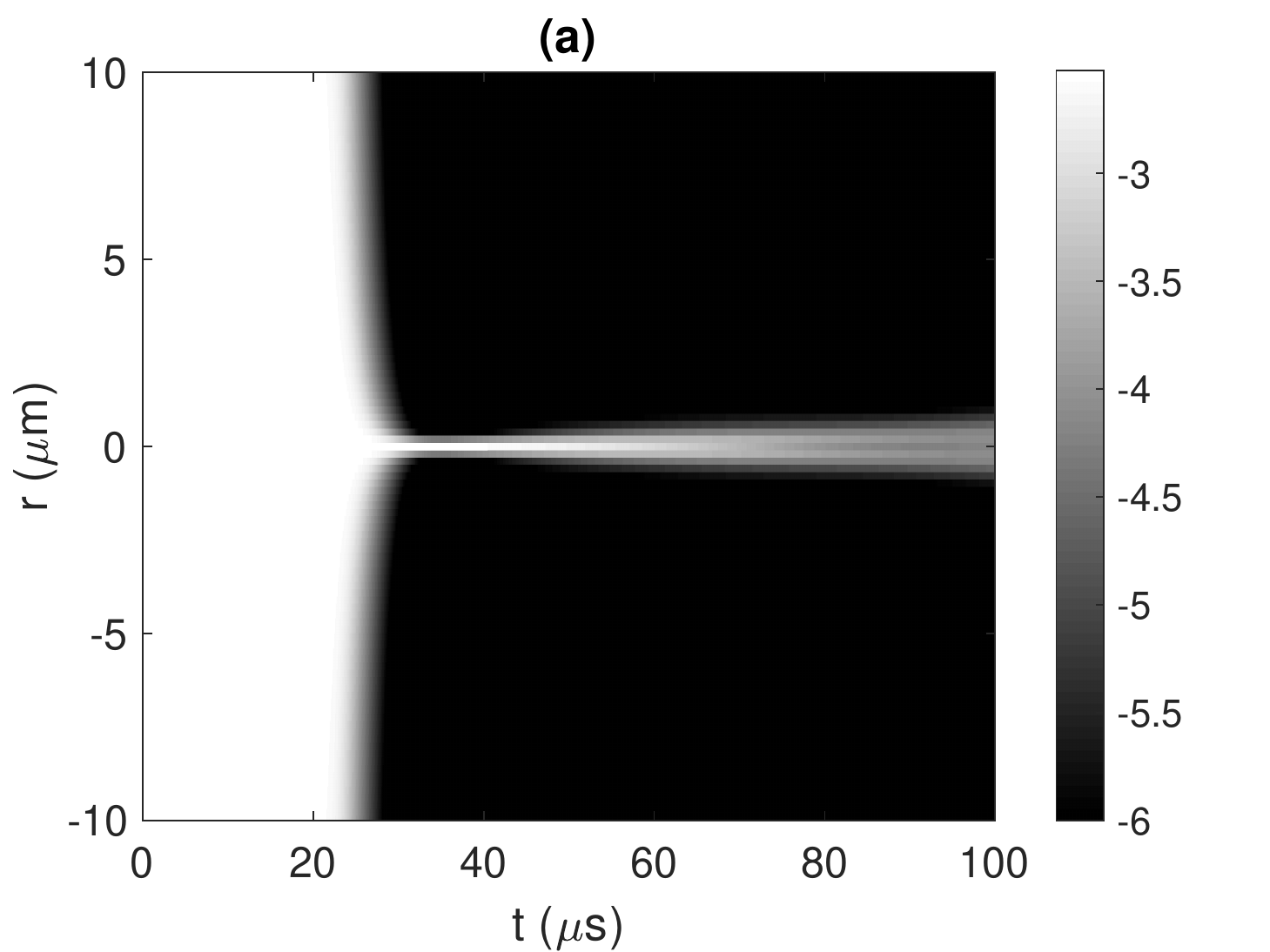}\includegraphics[width=0.5\columnwidth]{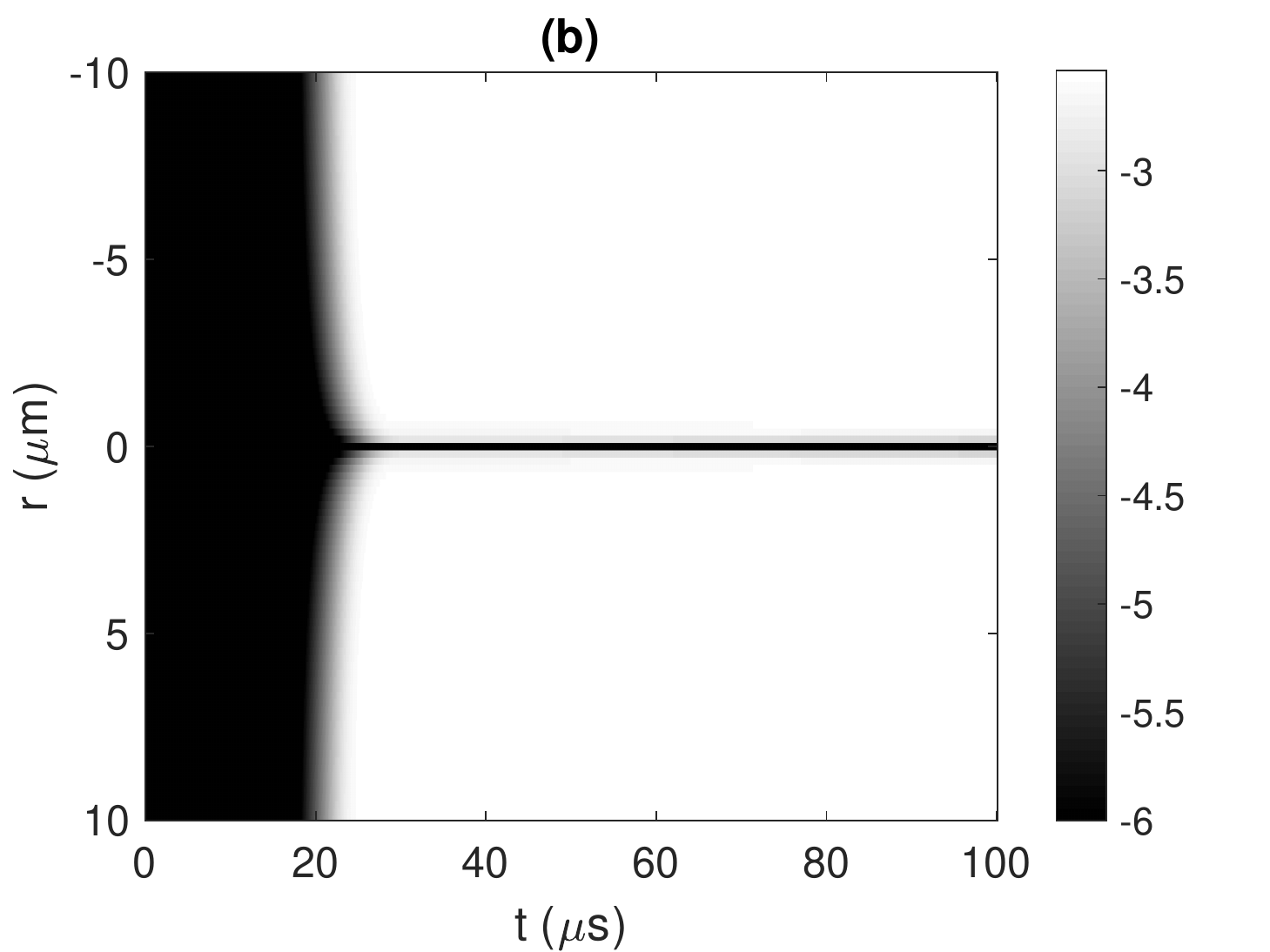}
\includegraphics[width=0.5\columnwidth]{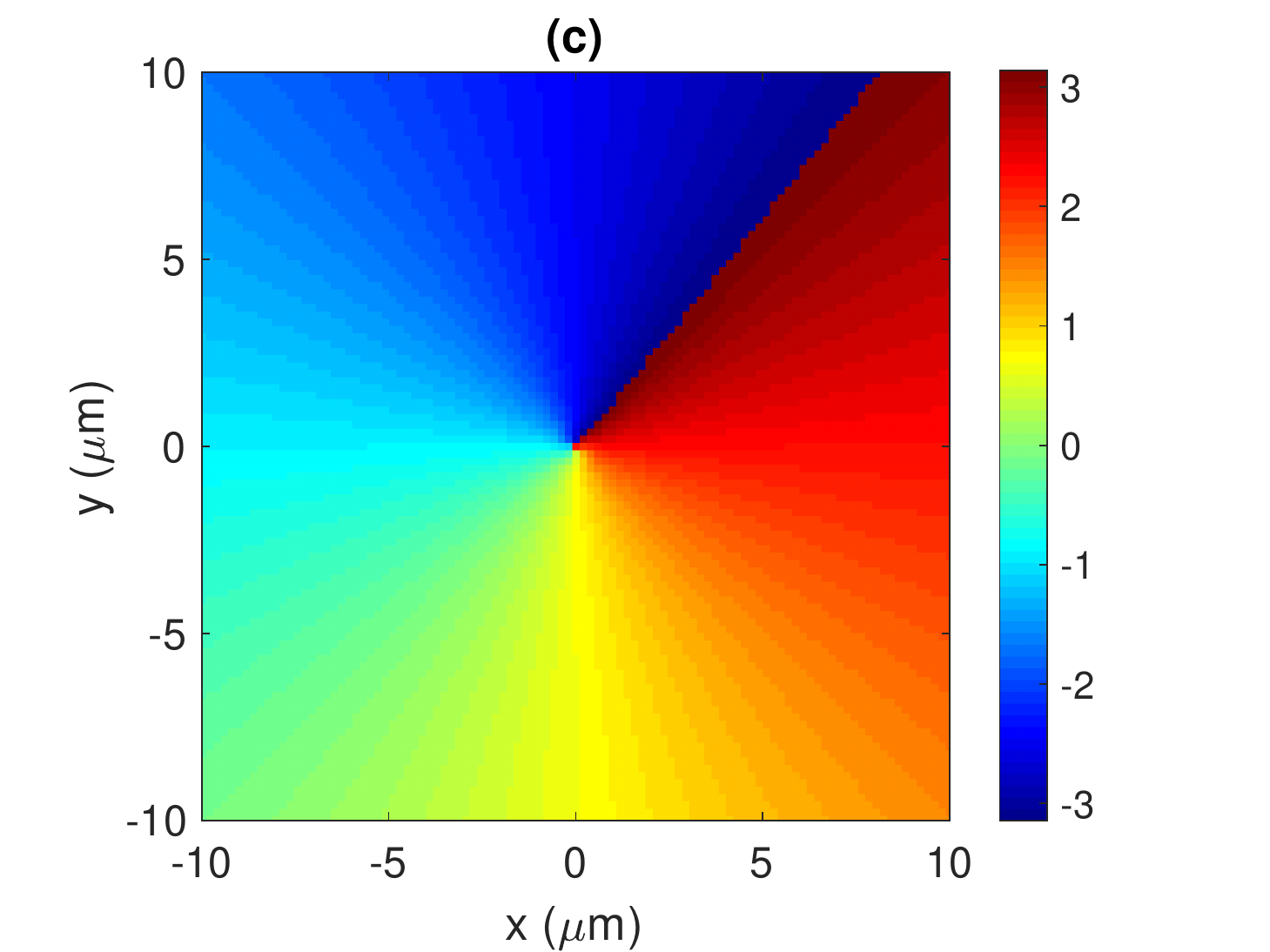}

\caption{Time evolution of densities $\log_{10}|\psi_{a}|^{2}$ (a) and $\log_{10}|\psi_{b}|^{2}$
(b) along the radial direction for the pump beam with OAM number $l=1$.
(c) Phase profile of the $b$ component at the end of time evolution.
The Stokes traveling wave $\Omega_{\mathrm{s}}(t)$ is given by Eqs.~(\ref{eq:1-1})
while the pump vortex beam $\Omega_{p}(t,r,\varphi)$ is described
by Eqs.~(\ref{eq:2-1}) and (\ref{eq:3-1}).}
\label{fig:fig7}
\end{figure}

Figure~\ref{fig:fig6} shows the full width at half maximum (FWHM)
of the density of component $a$ as it evolves in time. Figures~\ref{fig:fig7}(a)
and (b) show the time evolution of densities $\log_{10}|\psi_{a}|^{2}$
and $\log_{10}|\psi_{b}|^{2}$ along the radial direction, zooming
in on the region from $-10\ \mu m$ to $10\ \mu m$ where a localized
density structure forms. One can see that after applying the pump
and Stokes pulses the component $a$ develops an extremely narrow
2D spot-like structure at the core of vortex representing a 2D bright
soliton that slowly disperses. A part of population which is transferred
to the component $b$ creates a dark 2D soliton (a vortex), the phase
profile of the vortex beam (Fig.~\ref{fig:fig7}(c)) being imprinted
to this component via the 2D STIRAP. Thus one can create stable two
component vector dark-bright solitons.

\begin{figure}
\centering\includegraphics[width=0.5\columnwidth]{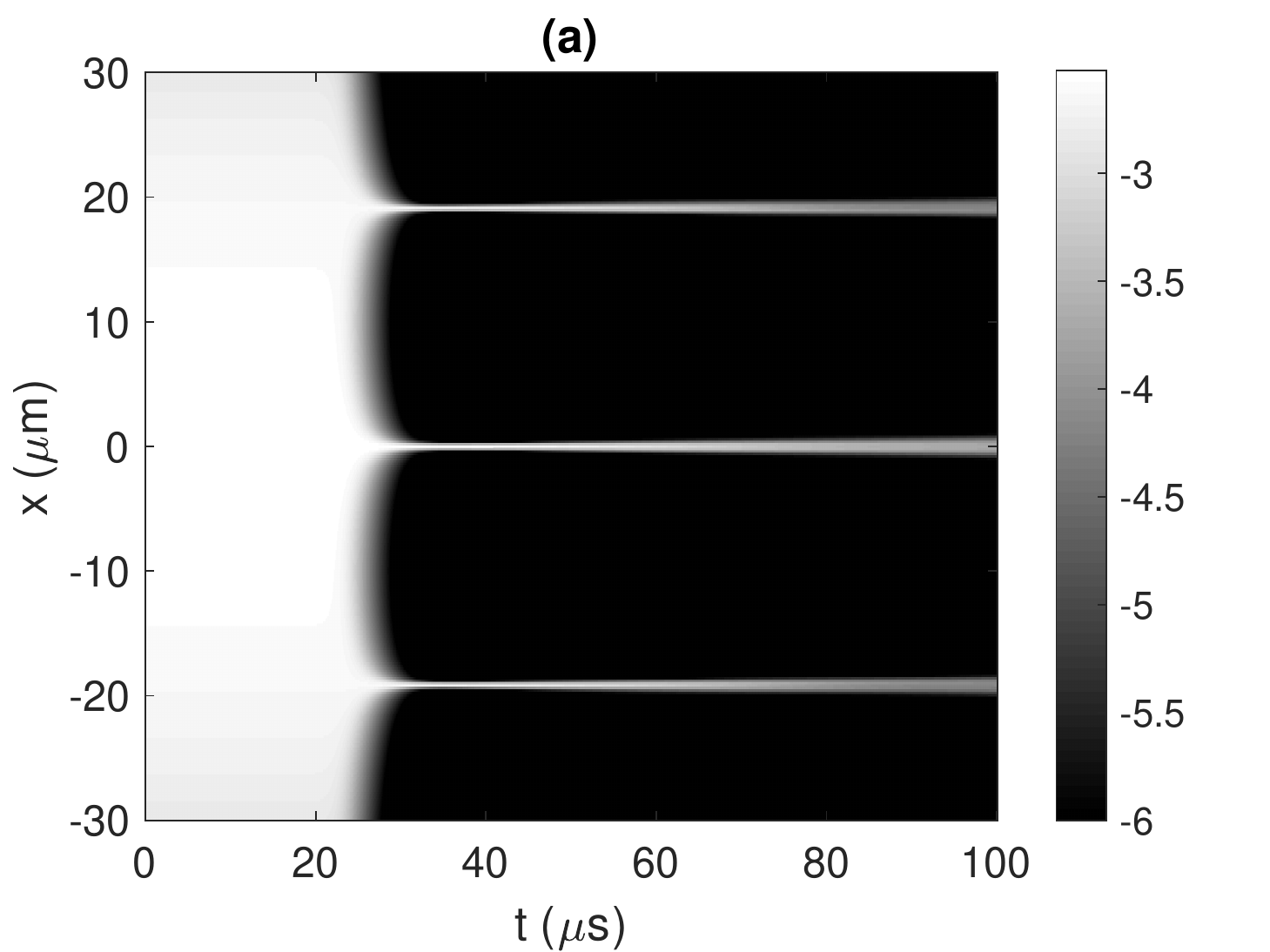}\includegraphics[width=0.5\columnwidth]{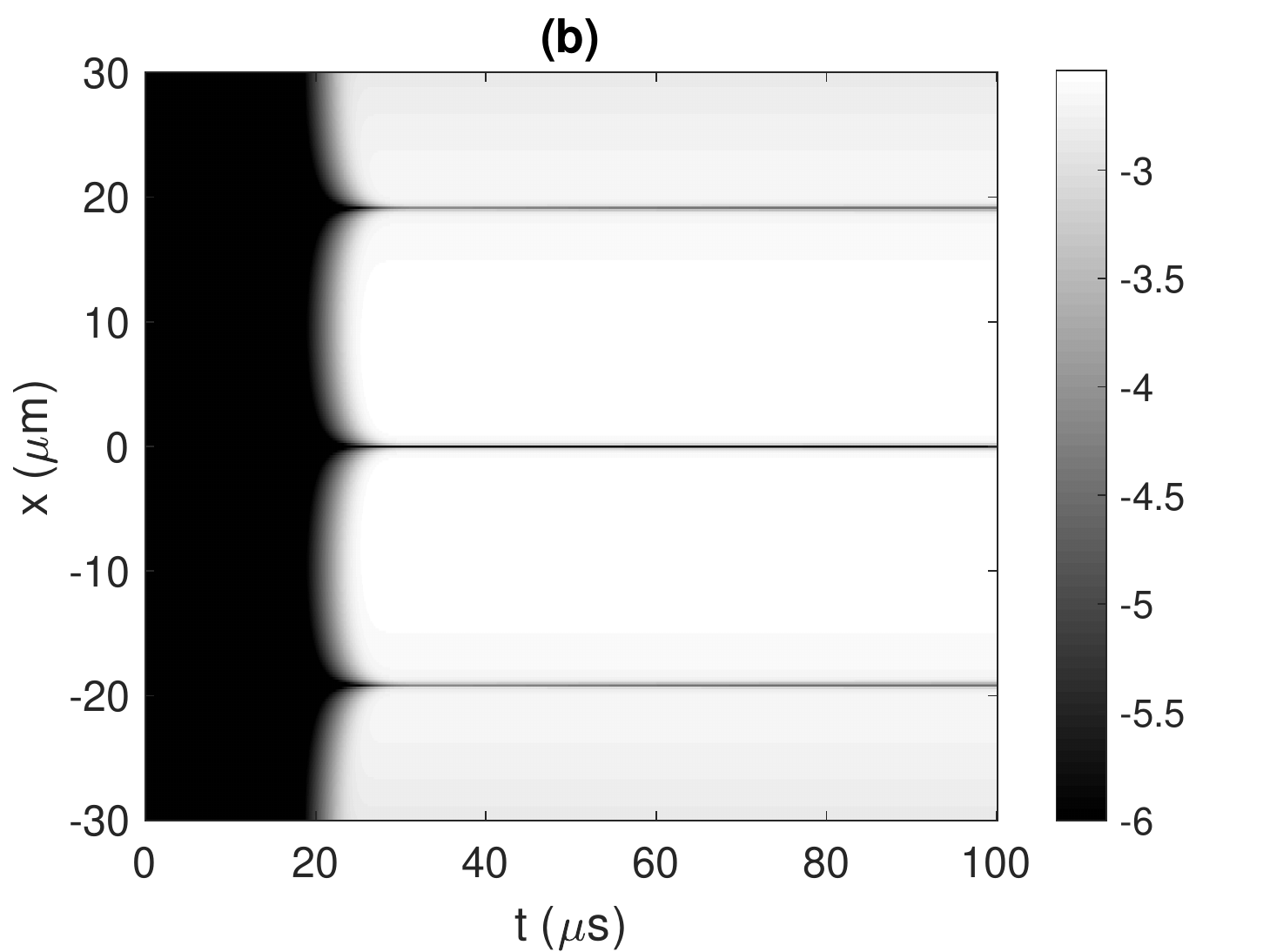}
\includegraphics[width=0.5\columnwidth]{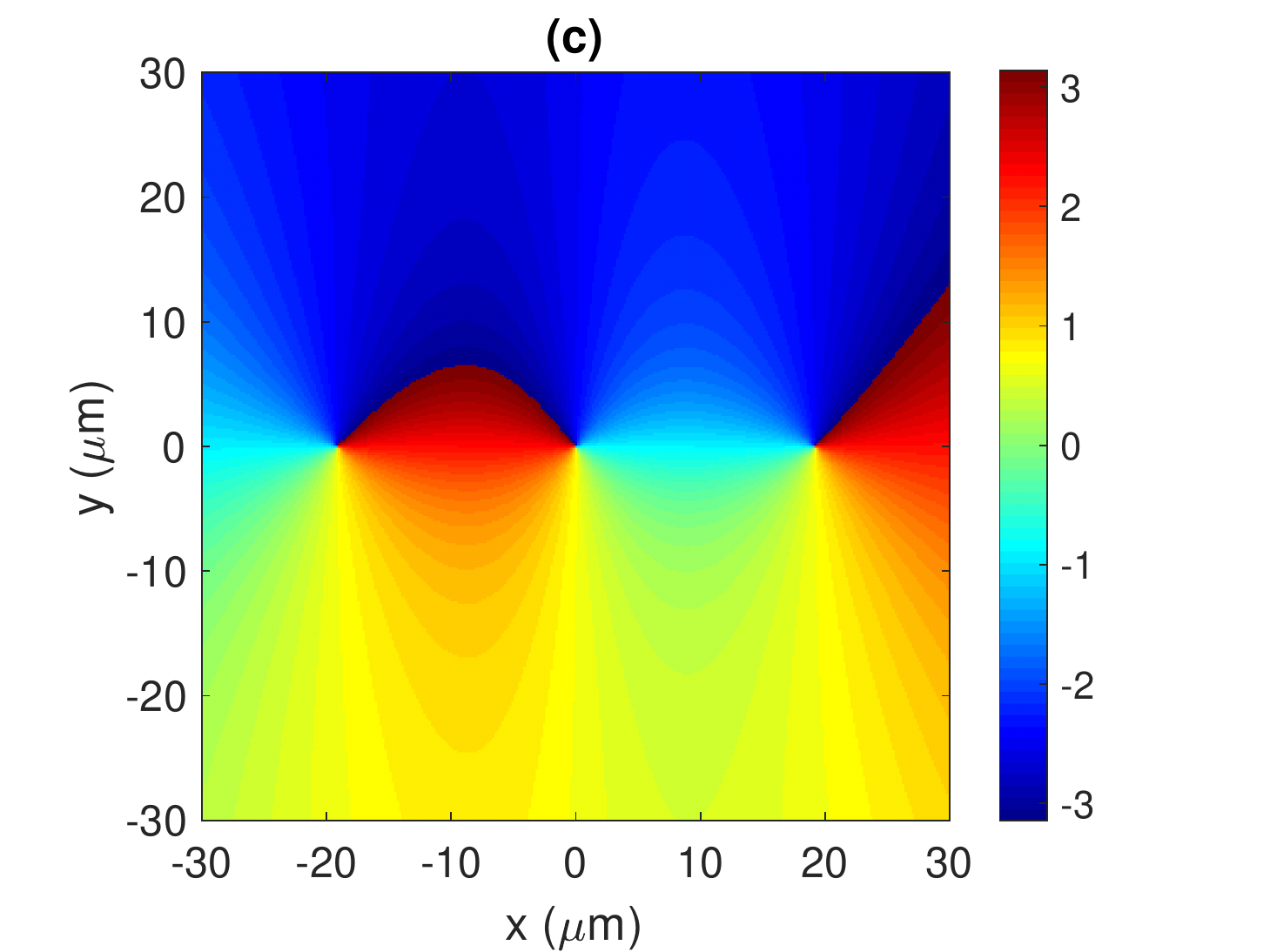}

\caption{Time evolution of densities $\log_{10}|\psi_{a}|^{2}$ (a) and $\log_{10}|\psi_{b}|^{2}$
(b) along the $x$ direction at $y=0$. (c) Phase profile of the $b$
component at the end of time evolution. The Stokes beam $\Omega_{\mathrm{s}}(t)$
is given by Eqs.~(\ref{eq:1-1}) while the pump beam $\Omega_{p}(t,r,\varphi)$
is described by Eqs.~(\ref{eq:2-1}) and (\ref{eq:composite beam}),
with $X=e^{i\pi}$, $r_{1}=\sqrt{(x+20)^{2}+y^{2}}$, $r_{2}=\sqrt{(x-20)^{2}+y^{2}}$,
$T=5\mu s$, $l_{1}=l_{2}=1$ and $w=20\mu m$.}
\label{fig:fig8}
\end{figure}

\begin{figure}
\centering\includegraphics[width=0.5\columnwidth]{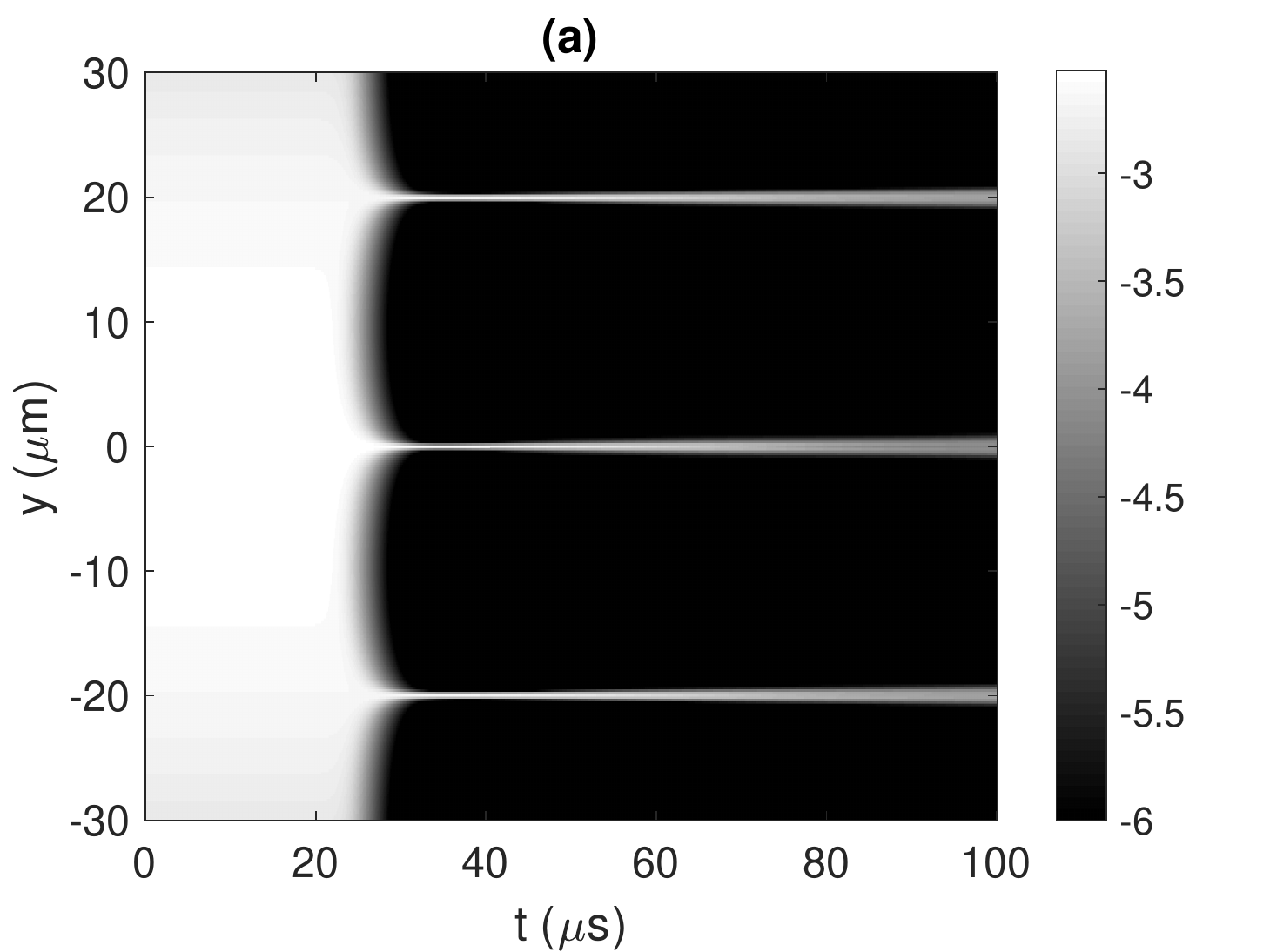}\includegraphics[width=0.5\columnwidth]{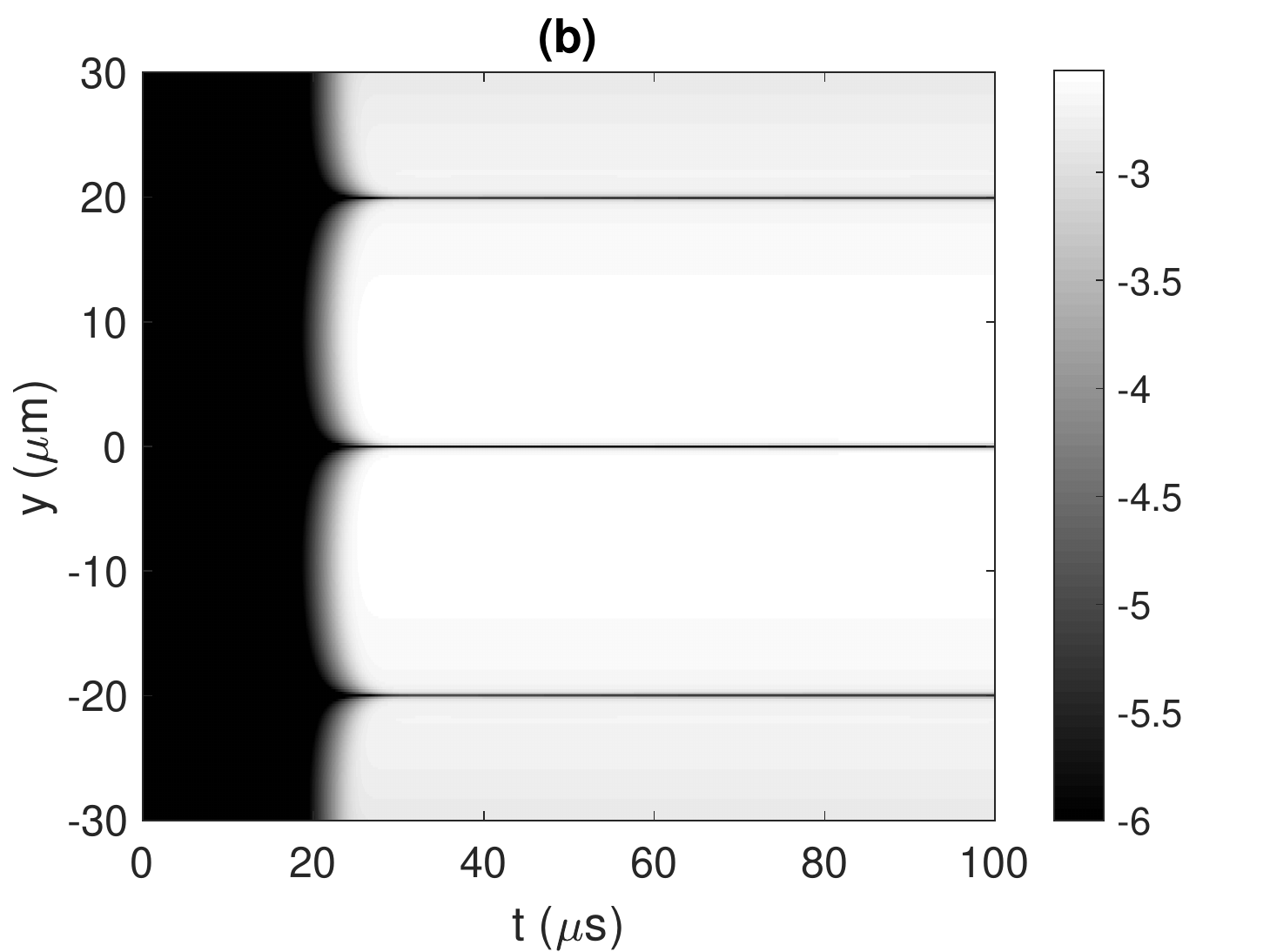}
\includegraphics[width=0.5\columnwidth]{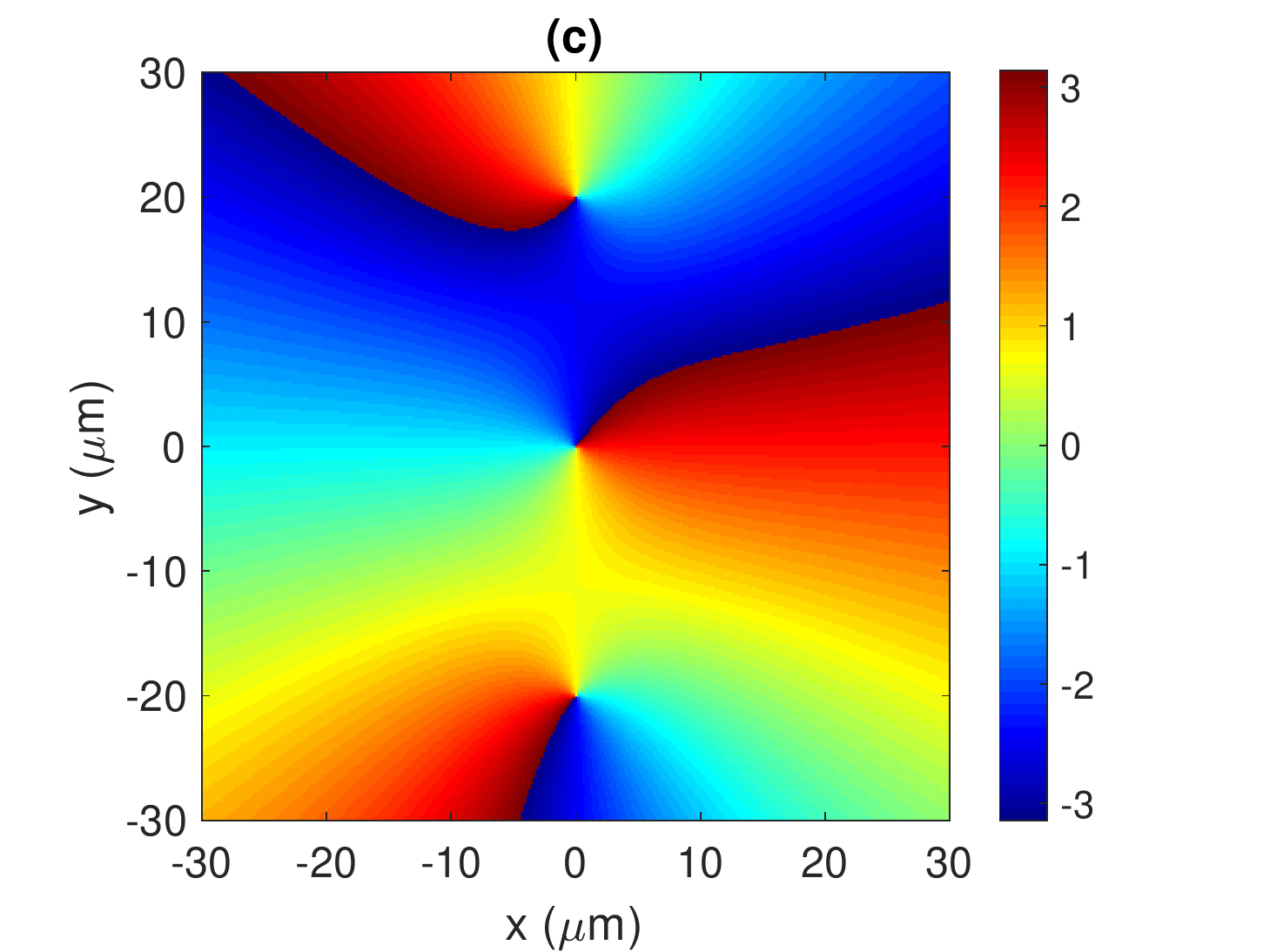}

\caption{Time evolution of densities $\log_{10}|\psi_{a}|^{2}$ (a) and $\log_{10}|\psi_{b}|^{2}$
(b) along the $y$ direction at $x=0$. (c) Phase profile of the $b$
component at the end of time evolution. The Stokes beam $\Omega_{\mathrm{s}}(t)$
is given by Eqs.~(\ref{eq:1-1}) while the pump beam $\Omega_{p}(t,r,\varphi)$
is described by Eqs.~(\ref{eq:2-1}) and (\ref{eq:composite beam})
with $X=1$, $r_{1}=r_{2}=r,$$\phi_{1}=\phi_{2}=\phi$, $l_{1}=1$
and $l_{2}=3$, the other parameters being the same as in Fig.~\ref{fig:fig8}.}
\label{fig:fig9}
\end{figure}

It is also possible to apply the 2D STIRAP protocol to the BEC by
using multiple vortices centered at different positions to create
several dark--bright defect structures with nanometer resolution.
In Fig.~\ref{fig:fig8}, we consider the pump beam of the form given
by Eq.~\ref{eq:composite beam} where $X=e^{i\pi}$, $r_{1}=\sqrt{(x+20)^{2}+y^{2}}$,
$r_{2}=\sqrt{(x-20)^{2}+y^{2}}$, $T=5\mu s$, $l_{1}=l_{2}=1$ and
$w=20\mu m$. On the other hand, if the beam shape is given by Eq.~(\ref{eq:composite beam})
with $X=1$, $r_{1}=r_{2}=r$ and $\phi_{1}=\phi_{2}=\phi$, the density
evolution and phase profiles are presented in Fig.~\ref{fig:fig9}
where $l_{1}=1$ and $l_{2}=3$. Clearly, several vortices are imprinted
to the BEC component $b$ along the $x$ or $y$ directions, creating
dark solitons. Multiple 2D structures with nanometer resolution are
also patterned in the component $a$ (bright solitons) at positions
where the 2D STIRAP process does not occur. The position of these
2D defects depends on the shape of original pump beam (\ref{eq:composite beam}).

Note that there exist some other proposals to transfer optical vortices
to BECs for creating the atomic defects \cite{NandiPhysRevA.69.063606,DuttonPhysRevLett.93.193602,KapalePhysRevLett.95.173601,AndersenPhysRevLett.97.170406,PugatchPhysRevLett.98.203601,WrightPhysRevA.77.041601,SimulaPhysRevA.77.015401,Ruseckas2011,Ruseckas2013,HouPhysRevA.96.011603}.
In particular, storing of the vortex slow light in the EIT configuration
allows to transfer a small amount of atomic population to the other
atomic internal ground state accompanied by the transfer of the vorticity
to the atoms \cite{DuttonPhysRevLett.93.193602,Ruseckas2011,Ruseckas2013}.
On the other hand, by applying the STIRAP with singular light beams
to atomic BECs one can create localized excitations such as vortices
\cite{NandiPhysRevA.69.063606} and solitons \cite{Gediminas2007}
with a more significant population transfer and without the diffraction
limitations occurring, e.g., in traditional ways of creating vortices
\cite{Dobrek1999,Lewenstein2001} and dark solitons \cite{Denschlag,BurgerPhysRevLett.83.5198,WuPhysRevLett.88.034101}
in BECs via the phase imprinting. In the method presented in ref.~\cite{NandiPhysRevA.69.063606},
the $\Lambda$-type BEC initially prepared in its ground level is
exposed to two copropagating laser pulses, viz. a nonvortex pump beam
and a vortex Stokes (dump) beam. This allows the transfer of angular
momentum from the Stokes field to the matter, creating a vortex BEC
in another atomic internal state. Yet, the method does not allow one
to create simultaneously a tightly localized 2D atomic population
in the initial atomic state at the vortex core of the Stokes beam,
as in this spatial region the atoms experience the absorption losses
by a non-vortex pump beam. Such a problem does not appear in the present
setup, as it is now the pump beam which carries an OAM with a zero
population at the vortex core. Thus one can create 2D two-component
dark-bright solitons described above. The present 2D STIRAP method
has also some similarities to the one considered in \cite{Gediminas2007}
for formation of two component (vector) solitons. Yet, ref.~\cite{Gediminas2007}
dealt with the 1D defects which are imprinted on the BEC via its interaction
with the first order Hermite--Gaussian mode and another two Gaussian
modes used in the tripod STIPAP process. On the other hand, the current
2D STIRAP method involving the Laguere-Gaussian mode permits generation
of the 2D defects in the BEC. 

The dark–bright defect structures proposed to generate in this paper can be measured via the standard time of flight and absorption imaging. Yet it is a destructive measurement. On the other hand, phase-contrast imaging, allowing the non-destructive measurement of condensate samples in situ \cite{AndrewsScience}, has proven to be a  useful technique to monitor vortices in Bose Einstein condensates. In particular, this technique has been used to observe vortices with the core filled by a second component \cite{anderson2000}, so it can also be applied to our proposal.

\section{Concluding Remarks \label{sec:conc}}

In summary, we have proposed a new method of 2D superlocalization
and patterning of atomic matter waves well beyond the diffraction
limit. For this we use two laser fields with a certain time delay,
an optical vortex pump as well as a Stokes pulses, interacting with
three internal levels of the matter wave. This configuration yields state-selective
strongly confined spot-like superlocalization patterns at the vortex
cores of the pump beams where the 2D STIRAP process does not take
place due to vanishing pump field. We have shown that the proposed
method can be implemented to imprint high contrast 2D defects with
nanometer resolution on an atomic BEC in a very controllable way.
Using this method one can also create stable multiple bright and dark
solitons close to each other with a full control over their size and
position. In this way, the present method allows to circumvent the
restriction set by the diffraction limit creating a variety of subwavelength
structures.

\section*{Appendix A: Population transfer efficiency}
The transfer efficiency should be sufficiently large, otherwise the
localization will wash out. The higher is the intensity, the better is the transfer.
From Fig. 2 of the manuscript, we see that the transfer efficiency is close to  $100\%$ outside the
region of zero intensity in the laser profile. This figure shows the remaining population after
pumping. Since the transfer is efficient, only a narrow density peak remains; otherwise we
would see a wider (or washed out) peak. This depends upon the adiabaticity criterion given
by Eq. 10. To clarify this point, we plot in Fig. 9 a 2D simulation of the spatial dependence of the adiabaticity criterion for the same parameters used
in Fig. 2(a). Note that the blue line shows the left hand side while the red line
corresponds to the right hand side in Eq. 10. The peak value of the criterion obviously shows that the
condition is always satisfied for our selected parameters, allowing a high transfer efficiency.
\begin{figure}
\centering\includegraphics[width=0.6\columnwidth]{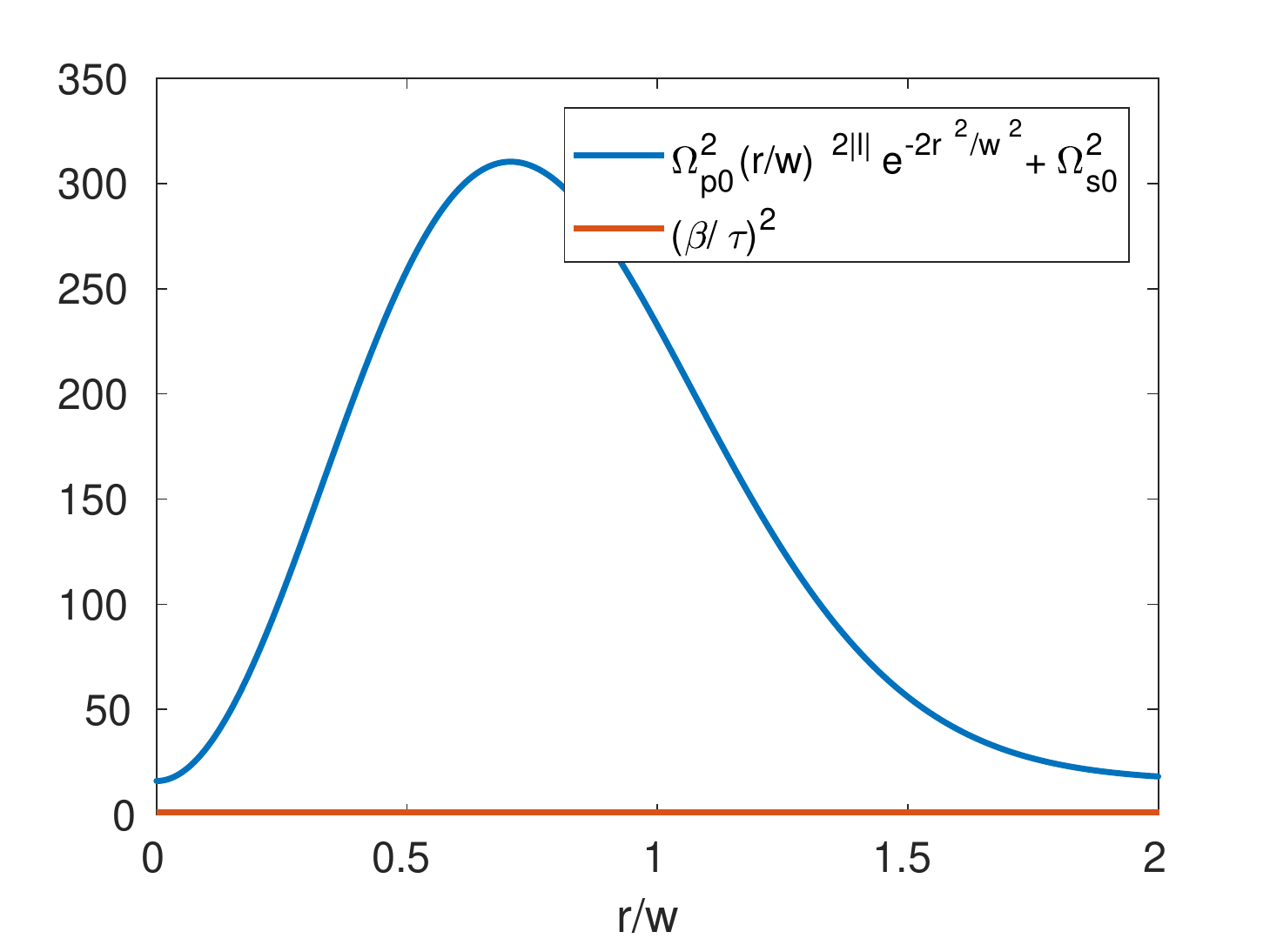}

\caption{A 2D numerical simulation of the spatial
dependence of the adiabaticity criterion given in Eq. 10 and for the parameters used in Fig. 2(a).}
\label{fig:fig10}
\end{figure}

\section*{Funding}
European Social Fund (09.3.3-LMT-K-712-19-0031); Ministerio de Econom\`ia y Competitividad, MINECO, (FIS2017- 86530-P and PID2020-118153GB-I00); Generalitat de Catalunya (SGR2017-1646); European Union Regional Development Fund within the ERDF Operational Program of Catalunya (project QUASICAT/QuantumCat);  European Cooperation in Science and
Technology (CA16221).

\section*{Acknowledgments}
This work has received funding from European Social Fund (Project
No. 09.3.3-LMT-K-712-19-0031) under grant agreement with the Research
Council of Lithuania (LMTLT) for H.R.H.
\\
\section*{Disclosures}
The authors declare no conflicts of interest.
\\
\section*{Data availability}
Data underlying the results presented in this paper are not publicly available at this time but may
be obtained from the authors upon reasonable request.
\\

%






\end{document}